\documentclass[twocolumn]{emulateapj}

\usepackage{natbib}
\usepackage{epsfig}
\usepackage{lscape}
\usepackage{graphicx}
\def\farcs{\hbox{$.\!\!^{\prime\prime}$}}  

\slugcomment{Submitted to ApJ on 15.10.2008, accepted on 27.12.2008}

\shorttitle{Radii and Temperatures of Exoplanet Host Stars}

\shortauthors{van Belle \& von Braun}

\begin{document}

\title{Directly Determined Linear Radii and Effective Temperatures\\of Exoplanet Host Stars}

\author{Gerard T. van Belle\altaffilmark{1}}
\author{Kaspar von Braun\altaffilmark{2}}

\altaffiltext{1}{European Southern Observatory,
Karl-Schwarzschild-Str. 2, 85748 Garching;
gerard.van.belle@eso.org}
\altaffiltext{2}{NASA Exoplanet Science Institute,
California Institute of Technology,
MC 100-22, Pasadena, CA 91125; kaspar@ipac.caltech.edu}

\begin{abstract}

We present interferometric angular sizes for 12 stars with known planetary companions, for comparison with 28 additional main-sequence stars not known to host planets.  For all objects we estimate bolometric fluxes and reddenings through spectral energy distribution fits, and in conjunction with the angular sizes, measurements of effective temperature.
The angular sizes of these stars are sufficiently small that the fundamental resolution limits of our primary instrument, the Palomar Testbed Interferometer, are investigated at the sub-milliarcsecond level and empirically established based upon known performance limits.
We demonstrate that the effective temperature scale as a function of dereddened $(V-K)_0$ color is statistically identical for stars with and without planets.
A useful byproduct of this investigation is a direct calibration of the $T_{\rm EFF}$ scale for solar-like stars, as a function of both spectral type and $(V-K)_0$ color, with an precision of $\overline{\Delta T}_{\rm {(V-K)}_0} = 138$K over the range $(V-K)_0=0.0-4.0$ and $\overline{\Delta T}_{\rm {SpType}} = 105$K for the range F6V -- G5V.  Additionally, we provide in an appendix spectral energy distribution fits for the 166 stars with known planets which have sufficient photometry available in the literature for such fits; this derived ``{\tt XO-Rad}'' database includes homogenous estimates of bolometric flux, reddening, and angular size.

\end{abstract}





\section{Introduction}\label{sec_introduction}

The formation, evolution, and environment of extrasolar planets are heavily influenced by their respective parent stars, including the location and extent of the habitable zone. To provide constraints on the characterization of these planets, it is therefore of significant scientific value to directly determine the astrophysical parameters of the host stars. Of particular interest are stellar radius $(R)$ and effective surface temperature $(T_{\rm EFF})$ since these two parameters help uniquely characterize our knowledge of extrasolar planet enviroments.  In the case of radius, planetary radii are frequently not directly measured but established through observations of transit events as a ratio of planet to stellar radius.  Measurements of planetary temperature are directly linked to the spectral characteristics of the star irradiating the planet.

For extrasolar planet hosting
stars (EHSs) that can be resolved with interferometers, their angular sizes $(\theta)$ are
directly measured.
Since the Stefan-Boltzman Law \citep{Stefan1879,Boltzmann1884} can be rewritten as
$T_{\rm EFF} \sim \left({F_{\rm BOL}} / {\theta^2}\right)^{
{1} / {4}}$,
\noindent
where $F_{\rm BOL}$ is the reddening-corrected bolometric flux, the effective temperature $T_{\rm EFF}$ can be directly measured for these stars.  We obtained data with the Palomar Testbed Interferometer (PTI) for 9 nearby EHSs with the aim of directly measuring their angular diameters, and computed estimates of their $F_{\rm BOL}$ through spectral energy distribution (SED) fitting to their available literature photometry. Additional EHS angular diameters from the Center for High Angular Resolution Astronomy (CHARA) Array are also folded into this investigation \citep{bvt07,baines2008ApJ...680..728B}. Further interferometric work relevant to the diameters of EHSs can be found in \citet{mjs91,mah03}.

Observational biases cause a large fraction of known EHSs to be nearby, enabling the use of Hipparcos parallaxes for a direct determination of their distances \citep{plk97, p97}; in combination with angular size measurements, their linear radii can be determined.\footnote{It is misleading, however, to indicate that interferometric angular size measurements independently lead to characterizations of stellar luminosity.  A {\it common mistake} is to assume that radius and temperature measurements derived from a single interferometric angular size can be combined through use of the Stefan-Boltzman law $(L \sim R^2 T_{\rm EFF}^4)$ to `measure' luminosity.  A cursory examination of the relationship between angular size and radius $(R \sim \theta)$ and temperature $(T_{\rm EFF} \sim  \theta^{-1/2})$ will demonstrate the new information contained in an angular size measurement is discarded when calculating $L$: only bolometric flux and distance information affect measures of $L$.}

The aim of this publication is to provide directly determined $R$ and $T_{\rm EFF}$  astrophysical parameters of these 12 EHSs along with equivalently derived parameters for a control group of 28 main sequence stars not currently known to host extrasolar planets. In addition, we present estimates of astrophysical parameters for all currently known EHSs with sufficient literature photometry (166 of the 230 known)\footnote{As of Feb. 1, 2008.}. The literature photometry and aforementioned SED fitting provides, for the sample of 166 EHS stars, estimates of $F_{\rm BOL}$ and $\theta_{\rm EST}$, done in the same way as done by \citet{vvc07}, if a $T_{\rm EFF}$ is assumed to be associated with the particular SED template being used to fit the stellar photometry.   These estimates of $F_{\rm BOL}$ and $\theta_{\rm EST}$ are presented in the ``{\tt XO-Rad}'' database at the end of this paper.

We describe the observations and data reduction of these stars in \S
\ref{sec_observations}; supporting data and spectral energy distribution fits are described in \S \ref{sec_sedfit}; derived effective temperatures and radii are presented in \S \ref{sec_parameters}, along with comparisons of our values to previous investigations (where available); finally, a detailed statical comparison of the EHS stars versus our control group is seen in \S \ref{sec_KStest}.


\section{Description of the Datasets}\label{sec_datasets}

We present interferometric results on two different datasets:

\begin{enumerate}
\item Known EHSs for which we were able to obtain PTI data
(\S \ref{sec_observations}) and calculate angular radii (\S
\ref{sec_parameters}). Knowledge of angular radii imposes an independent
constraint on the SED fitting (\S \ref{sec_sedfit}) and allows $T_{\rm EFF}$ to be
a determined directly. We were also able to augment our PTI data with 7 stars published from the CHARA Array by \citet{baines2008ApJ...680..728B}.  This dataset comprises 12 stars, 4 of which have data from both CHARA and PTI.
Together this sample of EHSs with angular sizes is our `EHSA' sample.
\item A number of main-sequence stars for which it was deemed possible to
resolve angular radii using PTI. This dataset comprises 28 stars and will be referred to as our `control sample'.
These stars are not currently known to host extrasolar planets and thus serve
as a comparison group for the EHSs with respect to astrophysical parameters.
\end{enumerate}
Additionally, SED fits are provided for all the well-characterized EHSs (status 1 Feb 2008, according to the Exoplanet
Encyclopedia\footnote{\url{http://exoplanet.eu/}}). The source dataset comprises
approximately 230 stars (including the ones for which we obtained PTI data),
although most of the fainter ($V>$10) stars are excluded due to a lack of available photometry;
they are presented in the ``{\tt XO-Rad}'' database in Appendix \ref{sec_appendix}. SED fitting for these
stars is performed based on literature
photometry and spectral templates with associated estimates of effective
temperatures.

\section{Supporting Data and Spectral Energy Distribution Fitting}
\label{sec_sedfit}

For all of the sources considered in this investigation, spectral
energy distribution (SED) fits were performed. Each fit, accomplished
using available photometry and an appropriate template spectrum,
produces estimates for the bolometric flux ($F_{\rm BOL}$), the angular
diameter ($\theta_{\rm EST}$) and the reddening ($A_{\rm V}$); effective
temperature during the SED fit is fixed for each of the template spectra.
In the absence of direct measurement of the angular diameter
(i.e. calibrators and stars listed in the {\tt XO-Rad} database), SED
fitting is used to estimate the angular size. When the angular
diameter is available from interferometrically measurements, SED fitting
is used to determine the bolometric flux and the reddening; effective
temperature as well as dereddened colors can then be derived.

These SED fits are
accomplished using photometry available in the literature as the
input values, with template spectra
from the \citet{p98} library appropriate for the spectral
types indicated for the stars in question. Spectral types used in the SED fitting for all EHS stars are those values found in the Exoplanet Encyclopedia, which is in turn based upon the respective source discovery papers cataloged therein.  The control sample stars as defined in \S \ref{sec_datasets} had their spectral types established from those values found in Hipparcos catalog \citep{plk97}.

The template spectra are adjusted by the fitting routine to account
for overall flux level, wavelength-dependent reddening, and expected
angular size.  Reddening corrections are based upon the empirical
reddening determination described by \citet{ccm89}, which differs
little from van de Hulst's theoretical reddening curve number 15
\citep{j68,dbv96}. Both narrowband and wideband photometry in the
0.3 $\mu$m to 30 $\mu$m are used as available, including Johnson
$UBV$ \citep[see, for example,][]{e63,m71}
Stromgren $ubvy\beta$
\citep{p76}, 2MASS $JHK_s$ \citep{csd03}, Geneva \citep{ruf76},
and Vilnius $UPXYZS$ \citep{zns72}; flux calibrations
are based upon the values given in \citet{Fukugita1995PASP..107..945F} and \citet{cox00}.
The results of the fitting for the calibrator stars is given in Table \ref{table_calibrators}; for the EHSA and control sample stars, Table \ref{tableData_LC5stars}, and for the ``{\tt XO-Rad}'' database, Table \ref{table_XOrad}.


\section{Observations and Data Reduction}\label{sec_observations}

\subsection{Visibility and Angular Sizes}\label{sec_vis_and_ang}

The calibration of the target star visibility $(V^2)$ data is performed by estimating the
interferometer system visibility ($V^2_{\textrm{\tiny SYS}}$) using the
calibration sources with model angular diameters and then normalizing the raw
target star visibility by $V^2_{\textrm{\tiny SYS}}$ to estimate the $V^2$
measured by an ideal interferometer at that epoch \citep{mjs91,bcv98,vv05}.
Uncertainties in the system visibility and the calibrated target visibility
are inferred from internal scatter among the data in an observation using
standard error-propagation calculations \citep{bkv99}. Calibrating our
point-like calibration objects against each other produced no evidence of
systematics, with all objects delivering reduced $V^2 = 1$.

Visibility and uniform disk angular size $(\theta_{\rm UD})$ are related using the first Bessel function $(J_1)$:  $V^2 =[2 J_1 (x) / x]^2$, where spatial frequency $x = \pi B \theta_{UD} \lambda^{-1}$. We may
establish uniform disk angular sizes for the target stars observed by the interferometer since
the accompanying parameters (projected telescope-to-telescope separation, or
baseline, $B$ and wavelength of observation $\lambda$) are well-characterized
during the observation.  The uniform disk angular size can (and should) be connected to a
more physical limb darkened angular size $(\theta_{\rm LD})$; however, this is a minor effect since $\theta_{\rm LD} / \theta_{\rm UD}$ is small  in the near-infrared \citep[$<1.5\%$; see,
for example,][]{st87,t94,dbv96,dvt98,dtb00}.


Strictly speaking, limb darkened angular size is utilized here as a reasonable proxy for the Rosseland angular diameter, which corresponds to the surface where the Rosseland mean optical depth equals unity, as advocated by \citet{st87} as the most appropriate surface for computing an effective temperature.  The dense, compact atmospheres of the stars considered in this investigation are well characterized by a uniform disk fit, and the small correction factors tabulated in \citet{dtb00} will be used to convert our $\theta_{\rm UD}$ sizes into the appropriate limb darkened $\theta_{\rm LD}$ numbers.  The number of visibility points $N(V^2)$, derived $\theta_{\rm UD}$ sizes, associated goodness-of-fit $\chi^2_\nu$ and residuals $(\delta V^2)$, \citet{dtb00} correction factors $\theta_{\rm LD} / \theta_{\rm UD}$ and resultant $\theta_{\rm LD}$ sizes are found in the first columns of Table \ref{tableData_LC5stars}.

\subsection{PTI Observations}\label{sec_PTIobservations}

PTI is an 85 to 110 m $H$- and $K$-band 1.6 $\mu$m and 2.2 $\mu$m)
interferometer located at Palomar Observatory in San Diego County, California,
and is described in detail in \citet{c99}.  It has three 40-cm apertures used
in pairwise combination for detection of stellar fringe visibility on
sources that range in angular size up to 5.0 milliarcseconds (mas), being
able to resolve individual sources with angular diameter $(\theta)$ greater than 0.60 mas in size.  PTI has been
in nightly operation since 1997, with minimum downtime throughout the
intervening years.  The data from PTI considered herein cover the range from
the beginning of 1998 (when the standardized data collection and pipeline
reduction went into place) until the beginning of 2008 (when the analysis of this
manuscript was begun).  In addition to the target stars discussed herein,
appropriate calibration sources were observed as well and can be found {\it en
masse} in \citet{vvc07}.  Additional calibration sources of minimal angular size, as discussed in \S \ref{sec_calibrationLimits}, were also selected and are listed in Table \ref{table_calibrators}.

\begin{deluxetable*}{lccrllccc}
\tablecolumns{9}
\tablewidth{0pc}
\tablecaption{New calibration sources used in this investigation, discussed in \S \ref{sec_PTIobservations}.\label{table_calibrators}}
\tablehead{
\colhead{Star} &
\colhead{RA} &
\colhead{DE} &
\colhead{$N_{\rm PHOT}$} &
\colhead{SpType} &
\colhead{Model} &
\colhead{$A_{\rm V}$} &
\colhead{$\chi^2_\nu$} &
\colhead{$\theta_{\rm EST}$}
}
\tabletypesize{\scriptsize}
\startdata
HD4058 & 00 43 28.09 & +47 01 28.7 & 75 & A5V & A5V & $0.000 \pm 0.007$ & 1.07 & $0.379 \pm 0.011$ \\
HD10205 & 01 40 34.80 & +40 34 37.6 & 55 & B8III & B5III & $0.236 \pm 0.010$ & 1.83 & $0.226 \pm 0.037$ \\
HD10874 & 01 47 48.00 & +46 13 47.6 & 21 & F6V & F6V & $0.000 \pm 0.015$ & 3.81 & $0.376 \pm 0.009$ \\
HD11529 & 01 56 00.00 & +68 41 07.0 & 39 & B8III & B5III & $0.213 \pm 0.013$ & 1.06 & $0.223 \pm 0.036$ \\
HD13476 & 02 13 41.61 & +58 33 38.1 & 78 & A3Iab & A2I & $1.614 \pm 0.012$ & 4.02 & $0.342 \pm 0.025$ \\
HD14212 & 02 19 16.85 & +47 22 48.0 & 37 & A1V & A0V & $0.000 \pm 0.011$ & 1.15 & $0.281 \pm 0.018$ \\
HD15138 & 02 27 51.75 & +50 34 12.7 & 26 & F4V & F2V & $0.392 \pm 0.013$ & 2.80 & $0.438 \pm 0.011$ \\
HD16399 & 02 38 00.70 & +07 41 43.4 & 66 & F6IV & F5IV & $0.064 \pm 0.011$ & 0.27 & $0.348 \pm 0.016$ \\
HD16582 & 02 39 28.95 & +00 19 42.7 & 79 & B2IV & B2IV & $0.133 \pm 0.014$ & 2.13 & $0.267 \pm 0.011$ \\
HD17163 & 02 45 20.87 & +04 42 42.2 & 66 & F0III: & F0III & $0.000 \pm 0.012$ & 0.58 & $0.287 \pm 0.021$ \\
HD18331 & 02 56 37.45 & -03 42 44.0 & 244 & A3Vn & A3V & $0.247 \pm 0.007$ & 3.53 & $0.386 \pm 0.013$ \\
HD20418 & 03 19 07.62 & +50 05 42.1 & 33 & B5V & B57V & $0.156 \pm 0.012$ & 2.43 & $0.233 \pm 0.050$ \\
HD23005 & 03 46 00.82 & +67 12 06.8 & 28 & F0IV & F02IV & $0.079 \pm 0.011$ & 0.24 & $0.396 \pm 0.023$ \\
HD23363 & 03 44 30.51 & -01 09 47.1 & 48 & B7V & B57V & $0.157 \pm 0.010$ & 0.91 & $0.205 \pm 0.043$ \\
HD24479 & 03 57 25.44 & +63 04 20.1 & 50 & B9.5V & B9V & $0.000 \pm 0.007$ & 1.79 & $0.298 \pm 0.045$ \\
HD35039 & 05 21 45.75 & +00 22 56.9 & 104 & B2IV-V & B2IV & $0.264 \pm 0.008$ & 3.59 & $0.208 \pm 0.008$ \\
HD36777 & 05 34 16.79 & +03 46 01.0 & 62 & A2V & A2V & $0.195 \pm 0.008$ & 4.54 & $0.340 \pm 0.015$ \\
HD37077 & 05 35 39.49 & -04 51 21.9 & 51 & F0III & F0III & $0.001 \pm 0.008$ & 0.94 & $0.416 \pm 0.030$ \\
HD41040 & 06 03 27.36 & +19 41 26.2 & 63 & B8V & B8V & $0.000 \pm 0.011$ & 3.28 & $0.246 \pm 0.042$ \\
HD42618 & 06 12 00.45 & +06 47 01.3 & 60 & G4V & G2V & $0.000 \pm 0.010$ & 2.30 & $0.380 \pm 0.007$ \\
HD46300 & 06 32 54.23 & +07 19 58.7 & 114 & A0Ib & A0I & $0.003 \pm 0.010$ & 3.85 & $0.375 \pm 0.019$ \\
HD86360 & 09 58 13.39 & +12 26 41.4 & 43 & B9IV & B6IV & $0.330 \pm 0.010$ & 3.43 & $0.261 \pm 0.013$ \\
HD89389 & 10 20 14.88 & +53 46 45.4 & 36 & F9V & F8V & $0.117 \pm 0.010$ & 0.50 & $0.420 \pm 0.007$ \\
HD91480 & 10 35 09.62 & +57 04 57.2 & 98 & F1V & F0V & $0.046 \pm 0.018$ & 0.43 & $0.499 \pm 0.014$ \\
HD93702 & 10 49 15.43 & +10 32 42.9 & 54 & A2V & A2V & $0.241 \pm 0.009$ & 4.08 & $0.359 \pm 0.016$ \\
HD96738 & 11 08 49.08 & +24 39 30.4 & 33 & A3IV & A0IV & $0.269 \pm 0.010$ & 1.71 & $0.257 \pm 0.015$ \\
HD97334 & 11 12 32.53 & +35 48 52.0 & 61 & G0V & G0V & $0.110 \pm 0.008$ & 0.24 & $0.460 \pm 0.008$ \\
HD97486 & 11 14 04.63 & +62 16 55.7 & 15 & G5III & G5III & $0.301 \pm 0.016$ & 1.33 & $0.354 \pm 0.022$ \\
HD102634 & 11 49 01.40 & +00 19 07.2 & 70 & F7V & F6V & $0.073 \pm 0.009$ & 0.71 & $0.426 \pm 0.010$ \\
HD103578 & 11 55 40.53 & +15 38 48.5 & 61 & A3V & A3V & $0.323 \pm 0.009$ & 3.09 & $0.335 \pm 0.012$ \\
HD104181 & 11 59 56.92 & +03 39 18.8 & 60 & A1V & A0V & $0.000 \pm 0.008$ & 1.72 & $0.276 \pm 0.017$ \\
HD106661 & 12 16 00.23 & +14 53 56.9 & 68 & A3V & A3V & $0.226 \pm 0.008$ & 1.35 & $0.395 \pm 0.014$ \\
HD110392 & 12 41 26.98 & +40 34 45.7 & 15 & K0III & K0III & $0.000 \pm 0.015$ & 5.20 & $0.389 \pm 0.021$ \\
HD111604 & 12 50 10.81 & +37 31 00.8 & 50 & A3V & A3V & $0.562 \pm 0.012$ & 2.62 & $0.324 \pm 0.012$ \\
HD113771 & 13 05 40.89 & +26 35 08.5 & 11 & K0III & K0III & $0.000 \pm 0.019$ & 3.69 & $0.419 \pm 0.023$ \\
HD114762 & 13 12 19.743 & +17 31 01.6 & 100 & F9V & F8V & $0.085 \pm 0.008$ & 1.88 & $0.286 \pm 0.005$ \\
HD119288 & 13 42 12.98 & +08 23 19.0 & 46 & F3Vp & F2V & $0.215 \pm 0.013$ & 2.29 & $0.396 \pm 0.010$ \\
HD119550 & 13 43 35.700 & +14 21 56.1 & 48 & G2V & G2V & $0.000 \pm 0.011$ & 0.22 & $0.372 \pm 0.007$ \\
HD119550 & 13 43 35.700 & +14 21 56.1 & 48 & G2V & G2V & $0.000 \pm 0.011$ & 0.22 & $0.372 \pm 0.007$ \\
HD119550 & 13 43 35.89 & +14 21 56.3 & 48 & G2V & G2V & $0.000 \pm 0.011$ & 0.22 & $0.372 \pm 0.007$ \\
HD121560 & 13 55 49.994 & +14 03 23.4 & 36 & F6V & F6V & $0.099 \pm 0.010$ & 0.79 & $0.441 \pm 0.010$ \\
HD125161 & 14 16 10.07 & +51 22 01.3 & 37 & A9V & A7V & $0.000 \pm 0.018$ & 1.41 & $0.468 \pm 0.012$ \\
HD128332 & 14 34 15.70 & +57 03 57.0 & 29 & F7V & F6V & $0.083 \pm 0.012$ & 1.19 & $0.381 \pm 0.009$ \\
HD140775 & 15 45 23.47 & +05 26 50.4 & 101 & A1V & A0V & $0.090 \pm 0.008$ & 0.92 & $0.263 \pm 0.017$ \\
HD141187 & 15 47 17.35 & +14 06 55.0 & 37 & A3V & A3V & $0.318 \pm 0.010$ & 5.94 & $0.311 \pm 0.011$ \\
HD142908 & 15 55 47.587 & +37 56 49.0 & 123 & F0IV & F02IV & $0.111 \pm 0.015$ & 0.24 & $0.480 \pm 0.028$ \\
HD144579 & 16 04 56.793 & +39 09 23.4 & 70 & G8V & G8V & $0.052 \pm 0.007$ & 2.09 & $0.509 \pm 0.010$ \\
HD144874 & 16 07 37.55 & +09 53 30.3 & 46 & A7V & A7V & $0.000 \pm 0.009$ & 0.91 & $0.311 \pm 0.008$ \\
HD150557 & 16 41 42.54 & +01 10 52.0 & 62 & F2.7III-IV & F2III & $0.000 \pm 0.011$ & 2.64 & $0.414 \pm 0.030$ \\
HD151900 & 16 50 22.25 & -02 39 15.3 & 58 & F1III-IV & F0III & $0.333 \pm 0.009$ & 4.87 & $0.304 \pm 0.022$ \\
HD158352 & 17 28 49.69 & +00 19 50.1 & 42 & A8V & A7V & $0.062 \pm 0.010$ & 1.18 & $0.357 \pm 0.009$ \\
HD164353 & 18 00 38.72 & +02 55 53.7 & 110 & B5Ib & B5I & $0.410 \pm 0.019$ & 1.53 & $0.436 \pm 0.018$ \\
HD164613 & 17 55 11.14 & +72 00 18.5 & 28 & F2.5II-III & F2III & $0.083 \pm 0.013$ & 3.31 & $0.497 \pm 0.037$ \\
HD169702 & 18 24 13.80 & +39 30 26.1 & 25 & A3IVn & A0IV & $0.217 \pm 0.014$ & 1.64 & $0.330 \pm 0.020$ \\
HD173649 & 18 44 48.19 & +37 35 40.4 & 76 & F0IVvar & F02IV & $0.000 \pm 0.012$ & 0.90 & $0.396 \pm 0.023$ \\
HD180482 & 19 16 31.02 & +04 50 05.4 & 46 & A3IV & A0IV & $0.380 \pm 0.012$ & 2.21 & $0.285 \pm 0.017$ \\
HD180777 & 19 09 09.75 & +76 33 38.9 & 67 & A9V & A7V & $0.234 \pm 0.011$ & 1.12 & $0.449 \pm 0.012$ \\
HD182564 & 19 20 40.07 & +65 42 51.9 & 38 & A2IIIs & A0III & $0.063 \pm 0.010$ & 1.81 & $0.427 \pm 0.062$ \\
HD184663 & 19 35 25.13 & +02 54 48.5 & 36 & F6IV & F5IV & $0.000 \pm 0.014$ & 1.14 & $0.339 \pm 0.016$ \\
HD186568 & 19 43 51.452 & +34 09 45.8 & 58 & B8III & B5III & $0.397 \pm 0.009$ & 3.94 & $0.144 \pm 0.023$ \\
HD192640 & 20 14 32.033 & +36 48 22.6 & 86 & A2V & A2V & $0.554 \pm 0.009$ & 2.44 & $0.495 \pm 0.023$ \\
HD198478 & 20 48 56.29 & +46 06 50.9 & 79 & B3Ia & B3I & $1.691 \pm 0.023$ & 0.62 & $0.416 \pm 0.056$ \\
HD199081 & 20 53 14.75 & +44 23 14.2 & 71 & B5V & B57V & $0.000 \pm 0.007$ & 2.76 & $0.238 \pm 0.051$ \\
HD200723 & 21 03 52.14 & +41 37 41.9 & 21 & F3IV & F02IV & $0.222 \pm 0.013$ & 0.64 & $0.328 \pm 0.019$ \\
HD202240 & 21 13 26.43 & +36 37 59.7 & 53 & F0III & F0III & $0.055 \pm 0.008$ & 4.96 & $0.290 \pm 0.021$ \\
HD210264 & 22 08 50.40 & +22 08 19.6 & 15 & G5III & G5III & $0.000 \pm 0.017$ & 1.33 & $0.414 \pm 0.026$ \\
HD214734 & 22 38 39.05 & +63 35 04.3 & 34 & A3IV & A0IV & $0.314 \pm 0.011$ & 2.21 & $0.325 \pm 0.020$ \\
HD217813 & 23 03 04.977 & +20 55 06.8 & 40 & G5V & G5V & $0.000 \pm 0.012$ & 5.09 & $0.431 \pm 0.008$ \\
HD218261 & 23 06 31.71 & +19 54 39.0 & 42 & F7V & F6V & $0.104 \pm 0.009$ & 1.96 & $0.387 \pm 0.009$ \\
HD218261 & 23 06 31.885 & +19 54 39.0 & 42 & F7V & F6V & $0.104 \pm 0.009$ & 1.96 & $0.387 \pm 0.009$ \\
HD218396 & 23 07 28.715 & +21 08 03.3 & 82 & A5V & A5V & $0.277 \pm 0.008$ & 3.21 & $0.282 \pm 0.008$ \\
HD218687 & 23 09 57.17 & +14 25 36.3 & 30 & G0V & G0V & $0.103 \pm 0.012$ & 0.71 & $0.436 \pm 0.008$ \\
HD220102 & 23 20 20.82 & +60 16 29.2 & 40 & F5II & F2II & $1.097 \pm 0.015$ & 1.07 & $0.426 \pm 0.032$ \\
HD220102 & 23 20 20.82 & +60 16 29.2 & 40 & F5II & F2II & $1.097 \pm 0.015$ & 1.07 & $0.426 \pm 0.032$ \\
HD223346 & 23 48 49.36 & +02 12 52.2 & 64 & F5III-IV & F5III & $0.042 \pm 0.011$ & 1.33 & $0.342 \pm 0.022$ \\
\enddata
\tablecomments{$N_{\rm PHOT}$ is the number of photometric data points available for the bolometric flux fitting;  SpType is the spectral type as reported by SIMBAD; Model is the spectral template chosen from \citet{p98} for the fitting; $\chi^2_\nu$ is the reduced chi-squared value of the fit, and $\theta_{\rm EST}$ is the estimated angular size from the fit.}
\end{deluxetable*}

\subsection{Limits of PTI Calibration}\label{sec_calibrationLimits}

As discussed by \citet{bcv98,bkv99}, PTI has an empirically established fundamental
limiting visibility measurement error of $\sigma_{V^2_{\textrm{\tiny
SYS}}}\approx 1.5$\%.
The source of this limiting night-to-night measurement error
is most likely a combination of
effects: uncharacterized atmospheric seeing (in particular, scintillation),
detector noise, and other instrumental effects.

This night-to-night
repeatability limit restricts the ultimate resolution of the instrument.  This
is at odds with the desire to measure stellar diameters which, for a given brightness, are quite
small in an angular sense relative to PTI's resolution.  Main sequence stars
are squarely in this regime for PTI, with only a few examples - those
considered in this investigation - that creep out of the nether regions of
point-like obscurity into the realm of resolvability.
Attempting to resolve stars at the edge of PTI's performance envelope requires careful consideration of the demonstrated limits of the
instrument, using the techniques
described in \citet[][henceforth Paper VB2]{vv05}.

For PTI, operating at the $K$-band with its 109-m N-S baseline, a target of 0.60 milliarcseconds (mas) in
size should have a normalized visibility of $V^2=94.89\%$ (as introduced in \S \ref{sec_vis_and_ang}).  As discussed in in
VB2, there is a strong motivation towards using calibration sources that are
as point-like as possible - generally speaking, one wishes to have calibration
sources that are significantly smaller than the targets being observed.  For
this investigation, to reach the regime of 0.60 mas targets, we restricted our
use of calibrators to those that are, on average, 0.35 mas or less in size.
These two size limits are selected to have sufficient numbers of sufficiently
bright targets and calibrators, respectively.

For such calibrators, observed by PTI, the visibility calibration limit
is $\sigma_{V^2}=0.186\%$ (from VB2, Equation 7), which
contributes an angular size error due to calibration of roughly 0.012 mas.
The night-to-night limiting $V^2$ measurement error of $\sigma_{V^2_{\textrm{\tiny
SYS}}}\approx 1.5\%$, however,
contributes an angular size error of 0.086 mas.  This is significant in that
the measurement error dominates any possible calibration bias, which is
particularly important when considering smaller targets.  If we were
instead to have selected calibrators closer to $\sim 0.70$ mas in size - more
typical of PTI investigations that observe larger targets that are $> 1$ mas in size -
then the calibration angular size error be $\sim 0.045$ mas, and would start
to compete with the measurement error in dominating the error budget.  This
would put our results at substantial risk of directly reporting any
measurement bias inherent in the process we used to estimate the angular sizes
of our calibration sources.  Since our goal is direct measurement of the
target angular sizes, we have taken great care to ensure that this is not the
case.

A second aspect of this consideration of PTI limiting performance is the
reported angular sizes of our target stars.  For stars that, after
calibration, report formal errors that are sufficiently small to be in
violation of PTI's known night-to-night repeatability, we increased their
reported angular size errors to the level consistent with that repeatability.
As a function of target angular size, we show the limits of angular size
accuracy possible with PTI's repeatability limit in Table \ref{tableCalibration}.
The first column shows various target angular sizes, followed by the
corresponding visibilities.  A calibrator of 0.35 mas, as noted above,
contributes a the limit on knowledge of visibility of $\sigma_{V^2}=0.186\%$;
the associated limit in angular size knowledge is then listed in column three.  The next two
columns list the night-to-night repeatability limit of $V^2$, and the
associated angular size error.  The final column combines the calibration
limit and the night-to-night limit in quadrature.

\begin{deluxetable*}{ccccccc}
\tablecolumns{7}
\tablewidth{0pc}
\tablecaption{Calibration floor by target angular size as discussed in \S \ref{sec_calibrationLimits}.\label{tableCalibration}}
\tablehead{
 &
&
\colhead{Calibration} &
\colhead{Calibration} &
\colhead{Night-to-night} &
\colhead{Night-to-night} &
\colhead{$\sigma_{\theta}$}
\\
\colhead{Target $\theta$}&
\colhead{Target $V^2$} &
\colhead{$\sigma_{V^2}$} &
\colhead{$\sigma_{\theta}$} &
\colhead{$\sigma_{V^2}$} &
\colhead{$\sigma_{\theta}$} &
\colhead{floor}
\\
\colhead{(mas)} &
&
&
\colhead{(mas)} &
&
\colhead{(mas)} &
\colhead{(mas)}
}
\tabletypesize{\scriptsize}
\startdata
0.600 & 0.94893 & 0.00186 & 0.012 & 0.01500 & 0.085 & 0.086 \\
0.650 & 0.94028 & 0.00186 & 0.010 & 0.01500 & 0.079 & 0.080 \\
0.700 & 0.93103 & 0.00186 & 0.010 & 0.01500 & 0.075 & 0.076 \\
0.750 & 0.92116 & 0.00186 & 0.010 & 0.01500 & 0.071 & 0.072 \\
0.800 & 0.91072 & 0.00186 & 0.009 & 0.01500 & 0.068 & 0.069 \\
0.850 & 0.89971 & 0.00186 & 0.008 & 0.01500 & 0.064 & 0.065 \\
0.900 & 0.88815 & 0.00186 & 0.008 & 0.01500 & 0.062 & 0.063 \\
0.950 & 0.87607 & 0.00186 & 0.008 & 0.01500 & 0.060 & 0.060 \\
1.000 & 0.86348 & 0.00186 & 0.008 & 0.01500 & 0.058 & 0.058 \\
\enddata
\end{deluxetable*}

\subsection{CHARA EHS Data}\label{sec_CHARAdata}

Additional angular diameters of EHSs were obtained with the Georgia State University CHARA Array \citep{baines2008ApJ...680..728B} with an intent of detecting possible face-on binarity masquerading as planetary companionship \citep{baines2008ApJ...682..577B}.  The CHARA Array is a optical/near-infrared interferometer similar to PTI \citep{tenbrummelaar2005ApJ...628..453T}, but with longer baselines (up to 330m), allowing for resolution of smaller objects.  For inclusion of the appropriate CHARA data into our dataset, we will apply observation criteria similar to the PTI data:  First, the calibration sources must be sufficiently unresolved, which we set for CHARA to be 0.50mas or less.  Second, the ratio of angular sizes of science targets and their calibrators must be greater than 1.5.  In applying these two criteria, we are confident that the resulting measured angular sizes are sufficiently independent of the calibrator angular sizes predicted by SED fitting.

The resulting dataset for inclusion in this analysis consists of seven EHS angular sizes from the CHARA investigation, of which four stars are common to both the PTI and CHARA samples (as noted \S \ref{sec_datasets}).  The ratios of the CHARA to PTI UD angular sizes for those four stars (HD3651, HD75732, HD143761, HD217014) are $1.15\pm 0.14, 1.05 \pm 0.10, 0.99\pm 0.13, 1.08 \pm 0.13$, respectively, with an overall weighted average ratio of $1.06 \pm 0.06$, indicating possibly a slight tendency for the PTI sizes to be too small (or the CHARA sizes to be too large), but this is a weak 1-$\sigma$ result.

As a further check on the consistency of the CHARA results and our techniques, we modeled the predicted SED sizes of the calibrators found in \citet{baines2008ApJ...680..728B}.  These results are seen in Table \ref{tab_bainesSEDs}; on average, our calibrator predictions are within $0.5 \sigma$, and no individual results are more than $1.9 \sigma$ away from \citet{baines2008ApJ...680..728B}.  Overall, we find that the CHARA and PTI results are excellent agreement with each other, despite independently developed methodologies.

\begin{deluxetable*}{cccccc}
\tablecolumns{6}
\tablewidth{0pc}
\tablecaption{Comparison of spectral energy distribution fits for calibrators from \citet{baines2008ApJ...680..728B} as discussed in \S \ref{sec_CHARAdata}.\label{tab_bainesSEDs}}
\tablehead{
\colhead{Target} &
\colhead{Calibrator} &
\colhead{Calibrator Size} &
\colhead{CHARA} &
\colhead{Difference} &
\colhead{$\sigma$} \\
\colhead{HD} &
\colhead{HD} &
\colhead{Est. (mas)} &
\colhead{Est. (mas)} &
\colhead{(mas)} &
\colhead{}
}
\startdata
3651 & 4568 & $0.363 \pm 0.008 $ & $0.347 \pm 0.006 $ & -0.016 & 1.6 \\
11964 & 13456 & $0.407 \pm 0.009 $ & $0.380 \pm 0.011 $ & -0.027 & 1.9 \\
19994 & 19411 & $0.484 \pm 0.030 $ & $0.485 \pm 0.019 $ & 0.001 & 0.0 \\
75732 & 72779 & $0.415 \pm 0.013 $ & $0.413 \pm 0.010 $ & -0.002 & 0.1 \\
143761 & 136849 & $0.236 \pm 0.035 $ & $0.255 \pm 0.016 $ & 0.019 & -0.5 \\
189733 & 190993 & $0.166 \pm 0.035 $ & $0.167 \pm 0.035 $ & 0.001 & 0.0 \\
217014 & 218261 & $0.387 \pm 0.009 $ & $0.384 \pm 0.015 $ & -0.003 & 0.2 \\
\enddata
\end{deluxetable*}

\begin{widetext}
\section{Stellar Parameters}\label{sec_parameters}

For both the EHSA and our control sample stars, the basic astrophysical parameters of effective temperature and linear radius are computed from the angular size data and ancillary supporting data.  These parameters are then compared between the two samples as a function of $(V-K)_0$ color and, in the case of temperature, spectral type; the results of sections \S \ref{sec_temperatures} and \S \ref{sec_radii} are found in Table \ref{tableResults_LC5stars}.


\subsection{Effective Temperatures}\label{sec_temperatures}

Stellar effective temperature, $T_{\rm EFF}$, is defined in terms of the star's luminosity and radius by $L = 4\pi \sigma R^2T_{\rm EFF}^4$. As noted in \S 1, rewriting this equation in terms of angular diameter $(\theta_{\rm LD})$ and bolometric flux $(F_{\rm BOL})$, $T_{\rm EFF}$ can be expressed as $T_{\rm EFF} = 2341 \times (F_{\rm BOL}/\theta^2_{\rm LD})^{1/4}$,
where $F_{\rm BOL}$ is in $10^{-8}$ ergs cm$^{-2}$ s$^{-1}$ and $\theta_{\rm LD}$ is in mas
\citep{vlt99}.  The derived temperature values for the resolved stars of this study are found in Table \ref{tableResults_LC5stars}, along with  $(V-K)_0$ color.
These temperatures are plotted versus $(V-K)_0$ in Figure \ref{fig_TvsVK}, and to explore any potential difference between the EHSA stars and the control sample, a fit of the  $T_{\rm EFF}$ versus $(V-K)_0$ trend is performed.

For the control sample, the initial fit reveals HD87901 as a significant outlier. This is most likely due to two factors: (1) HD87901 is bluest and hottest star, at $(V-K)_0=-0.352$ and $T_{\rm EFF}=14231 \pm 314$K, and (2) HD87901 is a rapid rotator with $v \sin i=300$ km/s \citep{abt2002ApJ...573..359A}, and will show departures from sphericity that induce gravity darkening which render individual $T_{\rm EFF}$ determinations meaningless \citep{Aufdenberg2006ApJ...645..664A}.  Omitting HD87901 from the fit, the best fit for the control sample stars is
\begin{equation}
T_{\rm EFF}=(2832 \pm 239) + (6511 \pm 225) \times 10^{(-0.2204 \pm 0.0255) \times (V-K)_0}
\end{equation}
with $\chi^2_\nu$=1.72, with the fitting and error ellipses following the techniques described in \citet{ptv92}.  (Inclusion of HD87901 in this fit returns $\chi^2_\nu$=4.98.)

If we include the EHSA stars in the fit, we find the CHARA data point for 55 Cnc (HD75732) a significant outlier as well, which we will discuss further in \S \ref{sec_75732}.  Omitting 55 Cnc from the unified fit, we find a single fit gives:
\begin{equation}\label{eqn_T_vs_VK}
T_{\rm EFF}=(2974 \pm 199) + (6368 \pm 208) \times 10^{(-0.2362 \pm 0.0227) \times (V-K)_0}
\end{equation}
with $\chi^2_\nu$=1.82.  This fit line is plotted in Figure \ref{fig_TvsVK}.
These fits indicate there is no statistically significant difference between the two populations (noting that the EHSA fit is poorly constrained with a small number of data points over a small range of $(V-K)_0$, preventing a fit to those data alone).
We revisit the question of population similarity in further detail in~\S\ref{sec_KStest}.

For the fit in Equation \ref{eqn_T_vs_VK}, the median value of the differences between the $T_{\rm EFF}$ values predicted by this fit and the measured $T_{\rm EFF}$ values is $\overline{\Delta T}_{\rm {(V-K)}_0} = 138$K.  Since the median value of the errors in the individual $T_{\rm EFF}$ measurements is $\overline{\sigma_T}=164$K, we believe the limit of precision in the line fit is not due to any intrinsic astrophysical scatter in the $T_{\rm EFF}$ versus $(V-K)_0$ relationship, but rather the limits of the current measurements.

Alternatively, a fit may be made for a cubic relationship between $T_{\rm EFF}$ and $(V-K)_0$,
\citep[see,
for example, the corresponding equation in ][]{lmo05}
but this produces no significant improvement:
\begin{equation}
T_{\rm EFF} = (9455 \pm 313) + (-3590 \pm 483) \times (V-K)_0 + (891 \pm 222) \times (V-K)_0^2
+ (-89 \pm 33) \times (V-K)_0^3
\end{equation}
with only $\chi^2_\nu$=1.68, in spite of the extra degree of freedom.

For those spectral types for which we have more than one stellar angular size measurement, we can compare the resultant weighted mean $T_{\rm EFF}$ values to the `canonical' values cited in \citet{cox00}, which can be traced back to the investigation by \citet{dejager1987A&A...177..217D}.  This comparison is seen in Table \ref{table_teff_sptype_cal}.  It is interesting to note that our values of $T_{\rm EFF}$ all track increasingly lower between types F8V to G2V in comparison to the \citet{dejager1987A&A...177..217D} values, before returning to agreement with those values at G5V and cooler.

Finally, given the large number of individual samples of our data set between types F6V and G5V, we present an empirical calibration of $T_{\rm EFF}$ versus spectral type for this full range, also in Table \ref{table_teff_sptype_cal}.  Spectral types that have no measurements (e.g., F7V) have $T_{\rm EFF}$ values interpolated from the adjoining spectral types. The average error by spectral type is $\overline{\Delta T}_{\rm SpType}=105$K. This table and Equation \ref{eqn_T_vs_VK} represent a direct calibration of the $T_{\rm EFF}$ scale for solar-like main sequence stars for the spectral type range F6V-G5V and color range $(V-K)_0=0.0 - 4.0$.  No attempt was made for $T_{\rm EFF}$ calibration for the later types due to the sparseness of the data, although our data at K1V, K7V and M2V represent $T_{\rm EFF}$ calibration for those specific spectral types.

\begin{deluxetable*}{cccccc}
\tablecolumns{6}
\tablewidth{0pc}
\tablecaption{Dereddened colors, effective temperatures and radii for luminosity class V stars, discussed in \S \ref{sec_parameters}.\label{tableResults_LC5stars}}
\tablehead{
\colhead{Star ID} &
\colhead{$V_0-K_0$} &
\colhead{$T_{\rm EFF}$} &
\colhead{$d$} &
\colhead{$R$} \\
&
\colhead{(mag)} &
\colhead{(K)} &
\colhead{(pc)} &
\colhead{$(R_\odot)$}
}
\tabletypesize{\scriptsize}
\startdata
\hline
\multicolumn{5}{l}{Control Sample: Stars not known to host planets:}\\
\hline
HD1326 & $4.095 \pm 0.053$ & $3584 \pm 105$ & $3.568 \pm 0.013 $ & $0.393 \pm 0.023 $ \\
HD4628 & $2.125 \pm 0.052$ & $4929 \pm 169$ & $7.460 \pm 0.048 $ & $0.749 \pm 0.051 $ \\
HD16160 & $2.247 \pm 0.052$ & $5262 \pm 216$ & $7.209 \pm 0.054 $ & $0.650 \pm 0.053 $ \\
HD16895 & $1.327 \pm 0.091$ & $6200 \pm 163$ & $11.232 \pm 0.100 $ & $1.313 \pm 0.069 $ \\
HD19373 & $1.395 \pm 0.071$ & $5722 \pm 110$ & $10.534 \pm 0.074 $ & $1.509 \pm 0.058 $ \\
HD20630 & $1.511 \pm 0.052$ & $5908 \pm 232$ & $9.159 \pm 0.065 $ & $0.882 \pm 0.069 $ \\
HD22484 & $1.358 \pm 0.101$ & $6618 \pm 449$ & $13.719 \pm 0.147 $ & $1.345 \pm 0.183 $ \\
HD30652 & $0.925 \pm 0.061$ & $7067 \pm 124$ & $8.026 \pm 0.061 $ & $1.217 \pm 0.043 $ \\
HD39587 & $1.404 \pm 0.071$ & $5766 \pm 144$ & $8.663 \pm 0.081 $ & $1.047 \pm 0.053 $ \\
HD87901 & $-0.352 \pm 0.061$ & $14231 \pm 314$ & $23.759 \pm 0.446 $ & $3.092 \pm 0.147 $ \\
HD88230 & $3.347 \pm 0.051$ & $4156 \pm 89$ & $4.873 \pm 0.019 $ & $0.649 \pm 0.028 $ \\
HD95735 & $4.031 \pm 0.051$ & $3593 \pm 60$ & $2.548 \pm 0.006 $ & $0.395 \pm 0.013 $ \\
HD97603 & $0.106 \pm 0.062$ & $8899 \pm 201$ & $17.693 \pm 0.260 $ & $2.281 \pm 0.106 $ \\
HD102647 & $0.194 \pm 0.052$ & $8759 \pm 158$ & $11.091 \pm 0.109 $ & $1.657 \pm 0.060 $ \\
HD109358 & $1.530 \pm 0.072$ & $5896 \pm 145$ & $8.371 \pm 0.058 $ & $1.025 \pm 0.050 $ \\
HD114710 & $1.311 \pm 0.100$ & $6167 \pm 165$ & $9.155 \pm 0.060 $ & $1.056 \pm 0.057 $ \\
HD119850 & $4.060 \pm 0.052$ & $3664 \pm 153$ & $5.431 \pm 0.037 $ & $0.481 \pm 0.040 $ \\
HD126660 & $1.175 \pm 0.073$ & $6358 \pm 161$ & $14.571 \pm 0.119 $ & $1.772 \pm 0.087 $ \\
HD141004 & $1.423 \pm 0.081$ & $6662 \pm 477$ & $11.754 \pm 0.111 $ & $1.060 \pm 0.152 $ \\
HD142860 & $1.168 \pm 0.062$ & $6496 \pm 153$ & $11.121 \pm 0.089 $ & $1.389 \pm 0.065 $ \\
HD149661 & $1.648 \pm 0.052$ & $5196 \pm 196$ & $9.778 \pm 0.081 $ & $0.934 \pm 0.070 $ \\
HD157881 & $3.419 \pm 0.052$ & $4030 \pm 242$ & $7.720 \pm 0.057 $ & $0.564 \pm 0.068 $ \\
HD185144 & $1.845 \pm 0.081$ & $5628 \pm 148$ & $5.767 \pm 0.015 $ & $0.678 \pm 0.035 $ \\
HD201091 & $2.546 \pm 0.051$ & $4526 \pm 66$ & $3.482 \pm 0.018 $ & $0.610 \pm 0.018 $ \\
HD201092 & $3.431 \pm 0.051$ & $4077 \pm 59$ & $3.503 \pm 0.009 $ & $0.628 \pm 0.017 $ \\
HD210027 & $1.267 \pm 0.071$ & $6359 \pm 141$ & $11.756 \pm 0.098 $ & $1.526 \pm 0.068 $ \\
HD215648 & $1.243 \pm 0.082$ & $6461 \pm 190$ & $16.250 \pm 0.203 $ & $1.787 \pm 0.106 $ \\
HD222368 & $1.245 \pm 0.081$ & $6521 \pm 179$ & $13.791 \pm 0.167 $ & $1.577 \pm 0.087 $ \\
\hline
\multicolumn{5}{l}{EHSA Sample: Known planet hosting stars (PTI):}\\
\hline
HD3651 & $1.914 \pm 0.051$ & $5438 \pm 324$ & $11.107 \pm 0.089 $ & $0.818 \pm 0.098 $ \\
HD9826 & $1.239 \pm 0.081$ & $6465 \pm 188$ & $13.468 \pm 0.131 $ & $1.480 \pm 0.087 $ \\
HD28305 & $2.168 \pm 0.052$ & $4990 \pm 50$ & $47.529 \pm 1.852 $ & $12.692 \pm 0.545 $ \\
HD75732 & $1.935 \pm 0.221$ & $4952 \pm 216$ & $12.531 \pm 0.132 $ & $1.100 \pm 0.096 $ \\
HD95128 & $1.180 \pm 0.341$ & $6140 \pm 294$ & $14.077 \pm 0.131 $ & $1.172 \pm 0.111 $ \\
HD117176 & $1.625 \pm 0.052$ & $5687 \pm 188$ & $18.109 \pm 0.239 $ & $1.858 \pm 0.124 $ \\
HD120136 & $0.933 \pm 0.053$ & $6680 \pm 260$ & $15.596 \pm 0.170 $ & $1.450 \pm 0.112 $ \\
HD143761 & $1.439 \pm 0.052$ & $5936 \pm 339$ & $17.428 \pm 0.216 $ & $1.306 \pm 0.149 $ \\
HD217014 & $1.432 \pm 0.051$ & $5800 \pm 338$ & $15.361 \pm 0.179 $ & $1.141 \pm 0.133 $ \\
\hline
\multicolumn{5}{l}{EHSA Sample: Known planet hosting stars (CHARA):}\\
\hline
HD3651 & $1.914 \pm 0.051$ & $5062 \pm 88$ & $11.107 \pm 0.089 $ & $0.944 \pm 0.033 $ \\
HD11964 & $1.543 \pm 0.022$ & $5413 \pm 359$ & $33.979 \pm 1.051 $ & $2.234 \pm 0.304 $ \\
HD19994 & $1.189 \pm 0.238$ & $6109 \pm 111$ & $22.376 \pm 0.376 $ & $1.898 \pm 0.070 $ \\
HD75732 & $1.831 \pm 0.042$ & $4836 \pm 75$ & $12.531 \pm 0.132 $ & $1.152 \pm 0.035 $ \\
HD143761 & $1.439 \pm 0.052$ & $5981 \pm 194$ & $17.428 \pm 0.216 $ & $1.287 \pm 0.084 $ \\
HD189733 & $2.051 \pm 0.028$ & $4939 \pm 158$ & $19.253 \pm 0.322 $ & $0.781 \pm 0.051 $ \\
HD217014 & $1.432 \pm 0.051$ & $5571 \pm 102$ & $15.361 \pm 0.179 $ & $1.237 \pm 0.047 $
\enddata
\end{deluxetable*}


\begin{figure*}
\plotone{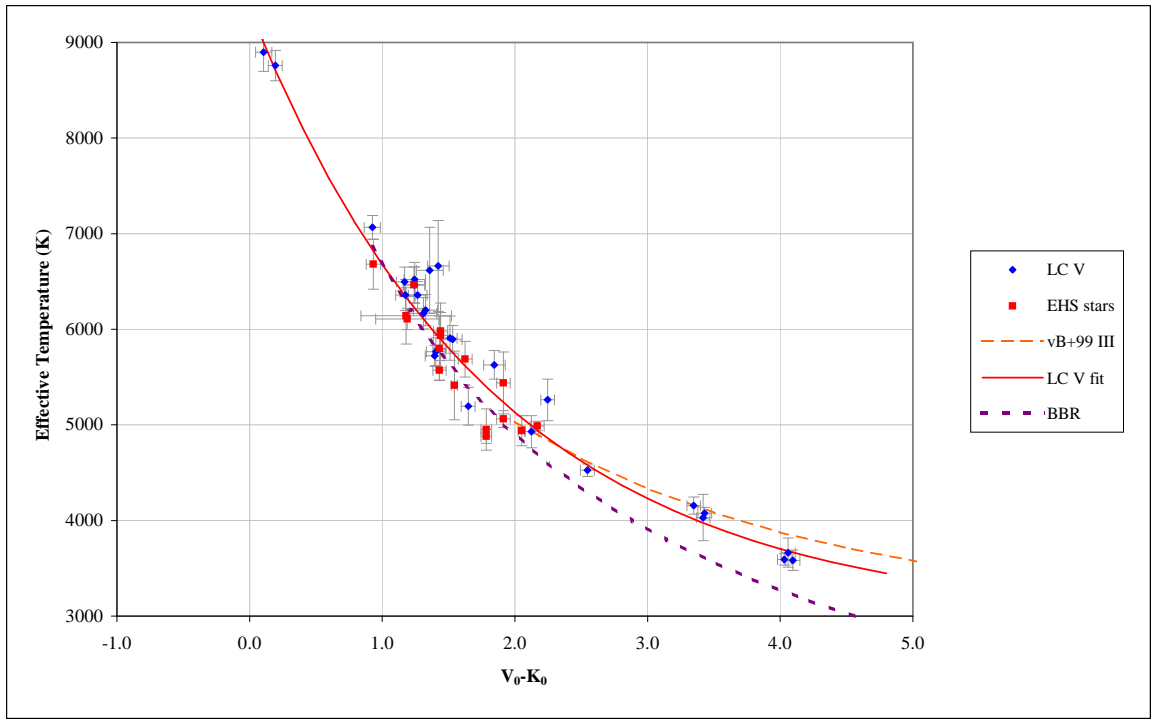}
\caption{\label{fig_TvsVK} Effective temperature $T_{\rm EFF}$ versus  $(V-K)_0$ color for control sample and EHSA stars.  Also shown is a fit to the luminosity class V stars (solid line, discussed in \S \ref{sec_temperatures}), the relationship for giants found in \citet{vlt99} (dashed line) and for a blackbody radiator (dotted line).  The median deviation of the stellar data points from the solid line fit is $\overline{\Delta T} = 138$K.}
\end{figure*}


\subsection{Linear Radii}\label{sec_radii}

From the parallax values found in Table \ref{tableData_LC5stars} from Hipparcos \citep{plk97}, linear radii are derived for the resolved stars of this investigation and are found in Table \ref{tableResults_LC5stars}.  A cubic relationship fit to the combined EHSA and control samples is:
\begin{equation}\label{eqn_R_VK}
R = (2.263 \pm 0.026) + (-1.261 \pm 0.016) \times (V-K)_0
+ (0.347 \pm 0.011) \times (V-K)_0^2
+ (-0.036 \pm 0.010) \times (V-K)_0^3
\end{equation}
with a $\chi^2_\nu$=15.1.  Clearly this metric indicates a poor fit, which is consistent with some of the stars beginning to evolve well off of the zero-age-main-sequence (ZAMS) line.  This effect is seen in a plot of the data in Figure \ref{fig_RvsVK}, with the presumably older stars being situated to the right of the line fit.  As such, Equation \ref{eqn_R_VK} should be regarded as only a rough indication of stellar radius, and not applicable in any general sense to determining linear radii of random field stars.

\end{widetext}

\begin{deluxetable}{ccccc}
\tablecolumns{5}
\tablewidth{0pc}
\tablecaption{Effective temperature versus spectral type, with an empirical calibration of effective temperature versus spectral type for types F6V through G5V. \label{table_teff_sptype_cal}}
\tablehead{
\colhead{Spectral} &
\colhead{N} &
\colhead{$T_{\rm EFF}$} &
\colhead{$T_{\rm EFF, Cox}$} \\
\colhead{Type} &
\colhead{} &
\colhead{(K)} &
\colhead{(K)}
}
\tabletypesize{\scriptsize}
\startdata
F6V & 6 & $6582 \pm 64$ & 6515\tablenotemark{a}\\
F7V &  & $6394 \pm 104$ & 6385\tablenotemark{a}\\
F8V & 4 & $6206 \pm 81$ & 6250\\
F9V &  & $6025 \pm 105$ & 6095\tablenotemark{a}\\
G0V & 7 & $5844 \pm 66$ & 5940\\
G1V &  & $5717 \pm 118$ & 5865\tablenotemark{a}\\
G2V & 2 & $5590 \pm 97$ & 5790\\
G3V &  & $5562 \pm 150$ & 5715\tablenotemark{a}\\
G4V &  & $5534 \pm 150$ & 5635\tablenotemark{a}\\
G5V & 4 & $5507 \pm 115$ & 5560\\
K1V & 4 & $4966 \pm 53$ & 4990\tablenotemark{a} \\
K7V & 3 & $4099 \pm 48$ &  4125\tablenotemark{a}\\
M2V & 3 & $3599 \pm 49$ & 3520\\

\enddata
\tablenotetext{a}{No specific value given in \citet{cox00}, interpolated from neighboring data points.}
\tablecomments{See discussion at end of \S \ref{sec_temperatures}.  Data after G5V were sufficiently sparse to not merit empirical calibration of the full range.  $N$ is the number of angular size measurements per spectral type;  rows with no value for $N$ are interpolated values. Columns 3 and 4 are from this work and \citet{cox00}, respectively.}
\end{deluxetable}

\begin{figure*}
\plotone{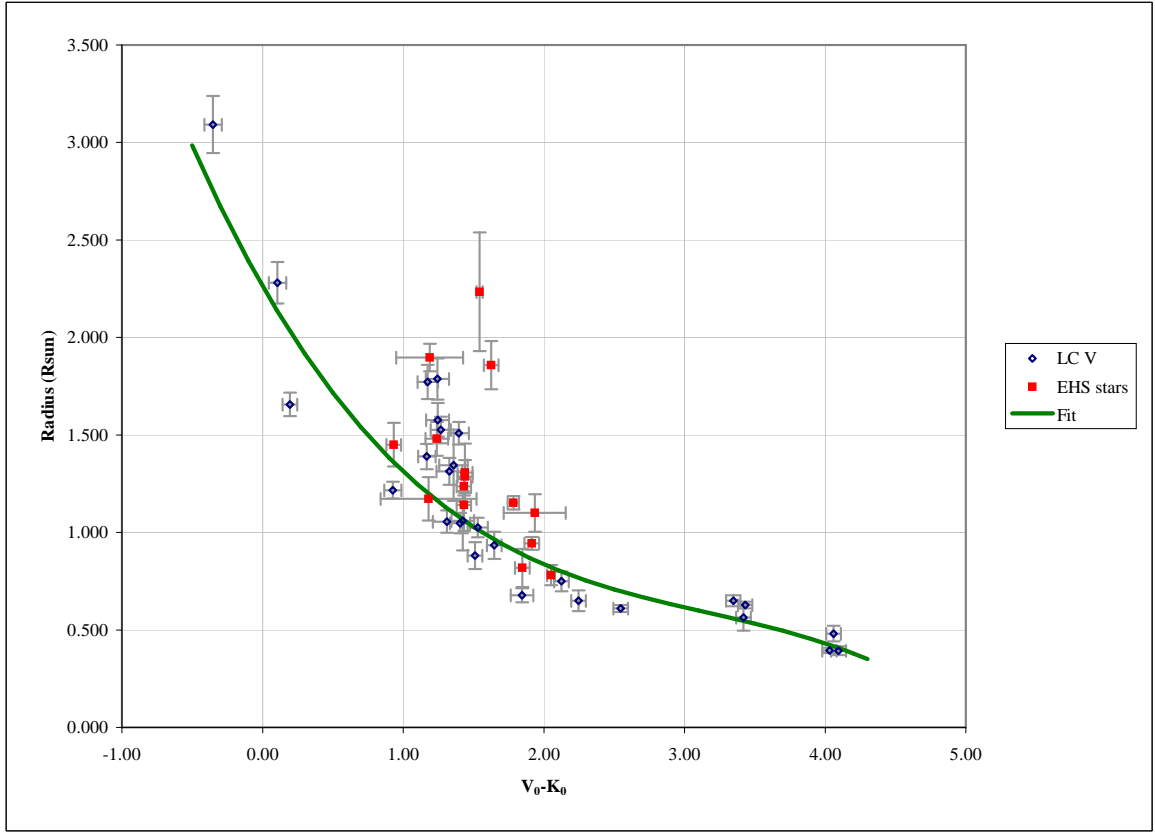}
\caption{\label{fig_RvsVK} Linear radius $R$ versus  $(V-K)_0$ color for control sample and EHSA stars as discussed in \S \ref{sec_radii}.  Also shown is a fit to the control sample and EHSA stars (solid line).  One of our EHSA stars, HD28305, is a giant star with $\{(V-K)_0=2.168 \pm 0.052, R=12.692 \pm 0.545 R_\odot \}$ and is off the scale of this plot.}
\end{figure*}



\subsection{Kolmogorov-Smirnov Comparison Between Exoplanet Hosting Stars and Control Stars} \label{sec_KStest}

As detailed in \citet{ptv92}, the Kolmogorov-Smirnov (KS) test can be executed to compare two arrays of data values, and examine the probability that the two arrays are drawn from the same distribution.  The KS test returns two values: the KS statistic $D$, which specifies the maximum deviation between the cumulative distribution of the two sample of data, and probability $p$, giving the significance of the KS statistic.  Small values of $p (<0.20)$ show that the two distributions differ significantly.

Examining the $T_{\rm EFF}$ versus $(V-K)_0$ data of the EHSA stars versus the control sample stars, we find that $D=0.25$ with $p=0.54$ -- strong indication that two data sets are indeed statistically indistinguishable.  The astrophysical implication is that, within the limits of our measurements, the effective temperature scale of stars with known planets does not differ from those without known planets.

The corresponding $R$ versus $(V-K)_0$  KS test, however, reports $D=0.50$ and $p=0.01$, which seems to indicate the two samples are inconsistent with each other.  However, the significance of this result is simple: our control sample is specifically selected to be main sequence stars, whereas the EHSA sample includes a number of evolved sources, as clearly seen in Figure \ref{fig_RvsVK}.  One corollary implication of these two KS tests is that stars on main sequence and those evolving off of it do not differ significantly in their $T_{\rm EFF}$ versus $(V-K)_0$ relationships.


\subsection{Comparison with Previous Studies}\label{secc2}

\begin{figure*}
\plotone{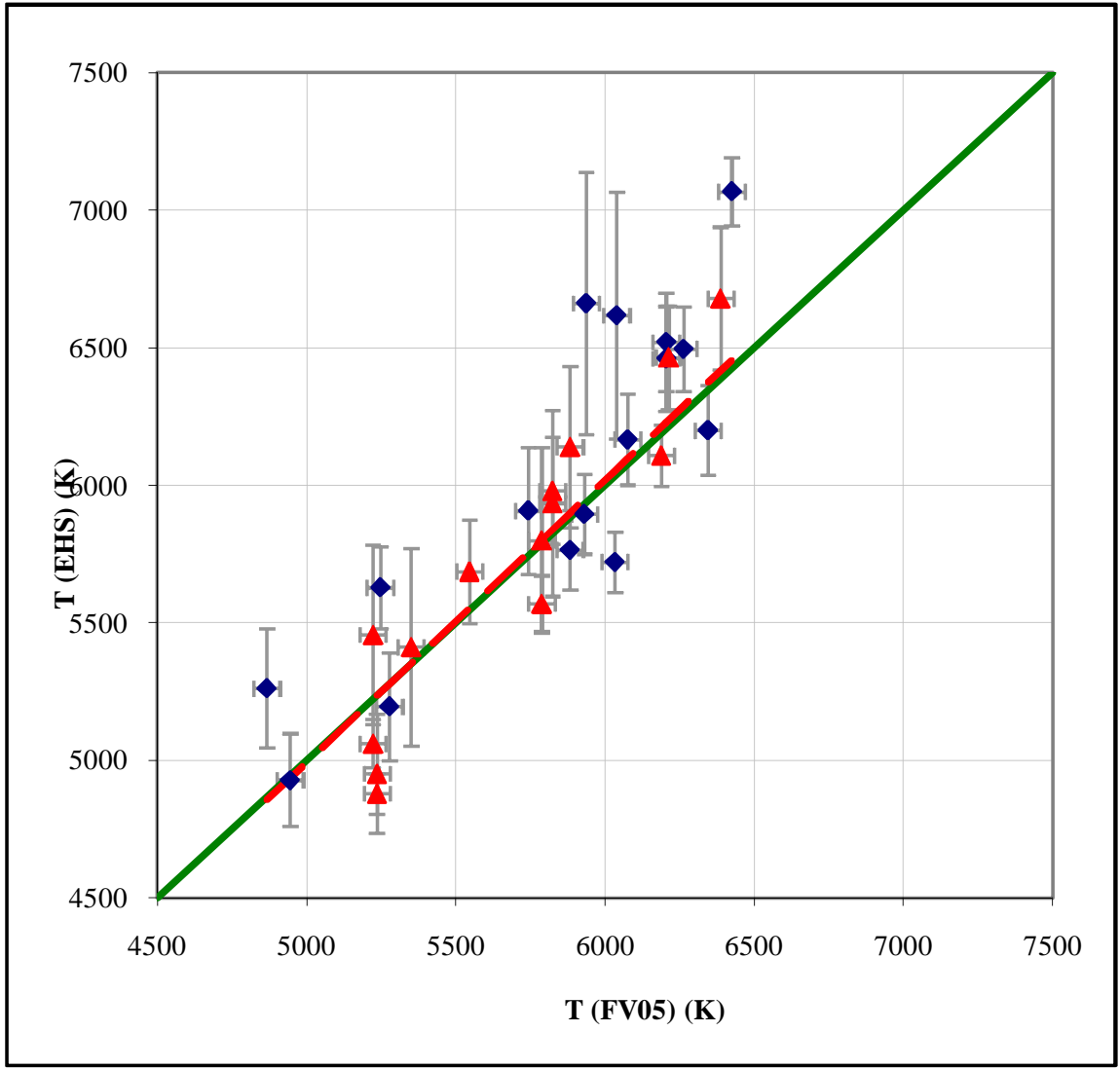}
\caption{\label{fig_TvsT} Effective temperature as determined by this study, versus those values found spectroscopically by \citet{vf05} for EHSA stars (red triangles) and our control sample stars (blue diamonds), as discussed in \S 5.3. The solid line is the 1:1 line, with the dotted line the fit to the $T_{\rm EHS} $ versus $ T_{\rm VF05}$ data points.}
\end{figure*}

There is a variety of data available for the known EHSs in the literature, derived from different methods by different authors. Thus, discrepancies, though sometimes small, exist. In order to be as consistent as possible, we chose the following two catalogs as data sources for astrophysical parameters:
\begin{itemize}
\item Mass, age, $T_{\rm EFF}$ and [Fe/H] from \citet{vf05}
\item Linear radius from \citet{tfs07}
\end{itemize} 
A comparison of the $T_{\rm EFF}$ values measured in this investigation can be directly contrasted against those found in \citet{vf05}.  Combining our EHSA and control star samples, we find:
%
\begin{equation}
T_{\rm EHS} = (-123 \pm 693) + (1.023 \pm 0.122) \times T_{\rm FV05}
\end{equation}
with $\chi^2_\nu=  1.66$.  As illustrated in Figure \ref{fig_TvsT}, there is no significant difference between the $T_{\rm EFF}$ values obtained with interferometry and spectroscopy.

\begin{figure*}
\plotone{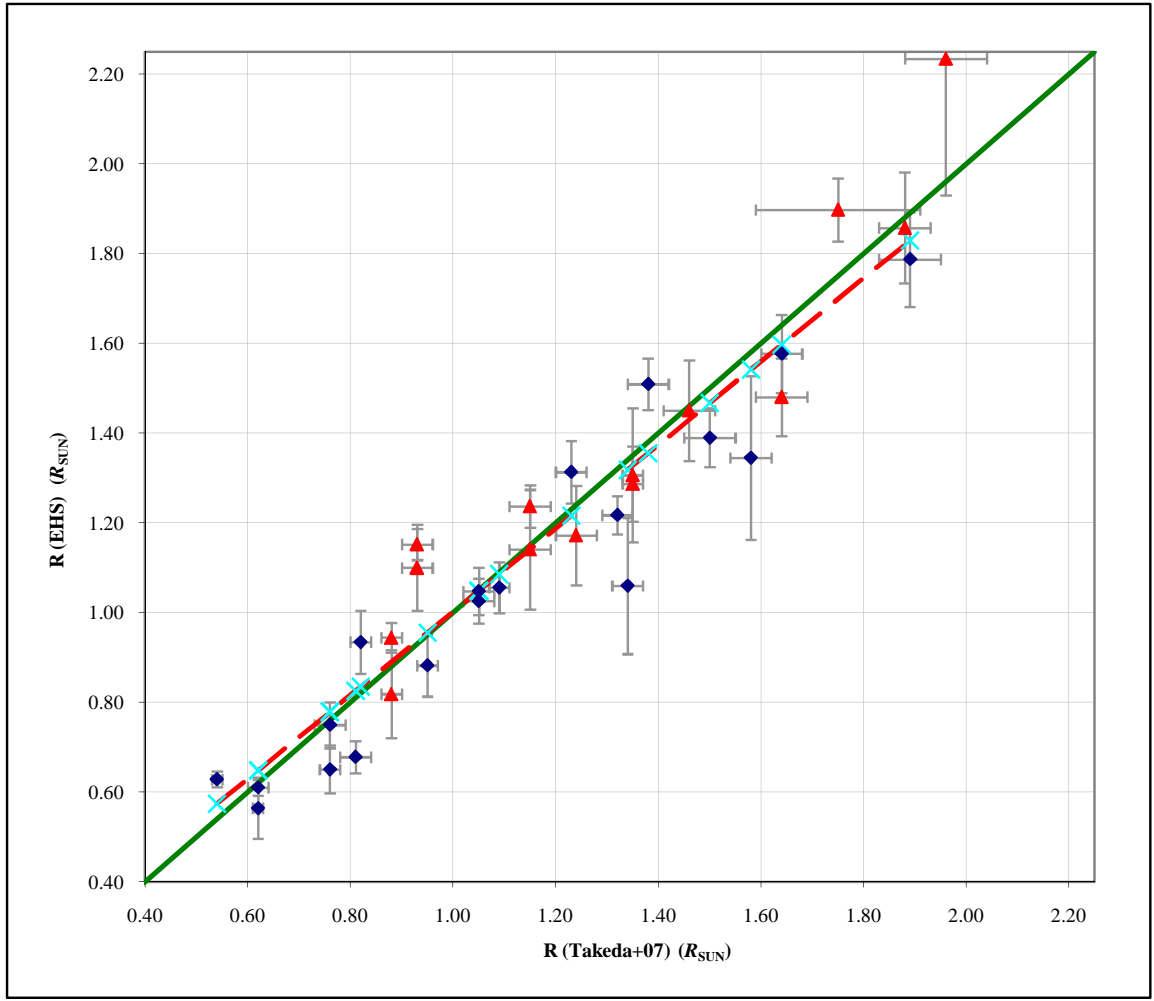}
\caption{\label{fig_RvsR} Linear radii as determined by this study, versus those values found spectroscopically by \citet{tfs07} for EHS stars (red triangles) and our control group (blue diamonds).  The solid line is the 1:1 line, with the dotted line the fit to the $R_{\rm EHS} / R_{\rm Takeda}$ values.  A trend is seen with the larger $(R>1.2 R_\odot)$ stars being larger in the \citet{tfs07} study.}
\end{figure*}

A marginal offset is found between our $R$ and the radii of \citet{tfs07}:
\begin{equation}
R_{\rm EHS} = (0.071 \pm 0.047) + (0.930 \pm 0.059) \times R_{\rm T+07}
\end{equation}
with $\chi^2_\nu=1.87$ -- roughly a 2-$\sigma$ offset between the line slope and intercept values for $R$ from theory versus those determined interferometrically.  The general trend is for the larger $(R>1.2 R_\odot)$ stars to have a larger theoretical, rather than interferometric, linear size. These values and the general trend can be seen in Figure \ref{fig_RvsR}.

\subsubsection{Discussion of 55 Cnc (HD 75732)}\label{sec_75732}


Inclusion of the PTI and CHARA data points for 55~Cnc\footnote{55 Cnc's distance is $12.53\pm0.13$ pc \citep{plk97}. It is K0IV--V star \citep{gcg03} with $V=5.398$ \citep{b00}. It has a mass of $0.92\pm0.046M_{\odot}$, an age of $9.5^{+3.4}_{-5.1}$ Gyr, $T_{\rm EFF}=5235\pm44$K, and [Fe/H]=$0.31\pm0.03$ \citep{vf05}. Its radius is 0.91 $R_{\odot}$ in \citet{ppc01} and $1.04\pm0.06 R_{\odot}$ when using the equations in \citet{l80}.  The values from this investigation are: $T_{\rm EFF}=4952\pm216$, $R=1.100\pm0.096  R_\odot$.} in the fit of Equation \ref{eqn_T_vs_VK} pushes the $\chi^2_\nu$ from 1.82 up to 2.91, with the CHARA data points remaining as $6-\sigma$ outliers; inclusion of just the PTI points results in $\chi^2_\nu$=1.88.  As such, we decided to omit both the CHARA and PTI data points for 55 Cnc from the fit.  There are two possible reasons for 55 Cnc turning up as ``too cool'' to fall onto the $T_{\rm EFF}$ versus $(V-K)_0$ fit of Equation \ref{eqn_T_vs_VK}.

First, the CHARA data points could be in error: including just the PTI data for 55 Cnc does not significantly alter the resulting $\chi^2_\nu$ value. However, the angular size and  $T_{\rm EFF}$ values for 55 Cnc from PTI and CHARA are in direct agreement with each other, although the PTI size data point has a larger error, indicative of its lesser resolving power for this $\sim 0.85$ mas star.  To `force' the 55 Cnc data onto the $T_{\rm EFF}$ versus $(V-K)_0$ fit line, its angular size would need to be reduced to $\sim0.70$mas.  Calibrator size error does not appear to be the source of the problem:  the size of calibrator HD72779 quoted in \citet{baines2008ApJ...680..728B} is $\theta_{\rm EST}=0.413\pm0.010$mas -- confirmed independently in this investigation with a value of $0.415\pm0.013$mas -- and would have to be $\sim 0.65$mas to push the 55 Cnc visibility data to deliver the larger angular size.  Alternatively, the $F_{\rm BOL}$ calculation for 55 Cnc could be too low, but require an increase from $1.4\times 10^{-8}$ erg~cm$^{-2}$~s$^{-1}$ to $\sim 2\times 10^{-8}$ erg~cm$^{-2}$~s$^{-1}$, which is far outside the allowable bounds of SED fitting, regardless of the template selected.

The second possible reason is that the visibility data could be contaminated by the presence of a secondary stellar companion.  Such a companion would reduce the observed visibility, resulting in an apparent increase in angular size, which in turn would effect an apparent decrease in derived temperature - as seen with the 55 Cnc data.  Examination of the $\{u,v\}$ plots associated with the CHARA dates and configuration cited in \citet{baines2008ApJ...680..728B} indicate a small amount ($<20^o$) of baseline rotation, with nearly zero change in baseline length, which would have lead to a null result in detection in \citet{baines2008ApJ...682..577B} for a secondary stellar companion - even in some cases where one is present.  There is a known companion to 55 Cnc at a distance of $\sim$1000 pc, or 9\farcs{}5 on the sky; however, with $\Delta K=3.65$ (based on a spectral type of $\sim$M4), in the worst case we would see a visibility change of  of only $\Delta V\sim 0.02$, which would only lower the apparent size from $\sim$0.85 to 0.82 mas.
Additionally, our na\"ive expectation is that the intensive spectroscopic studies of 55 Cnc that have turned up no less than 5 planets \citep{Fischer2008ApJ...675..790F} would have uncovered such a companion, so we are at a loss as to how to reconcile interferometric data with the spectroscopic discoveries.  For the moment we will be content to simply remove it from the effective temperature scale calibrations presented in \S \ref{sec_temperatures}.

\section{Summary and Conclusion}\label{sec_conclusion}

We present directly determined stellar radii and effective temperatures for 12 exoplanet host stars, along with the same estimates for 28 main-sequence control stars not known to host planets. In the process, we demonstrate the empirical limit of PTI's stellar angular resolution and the implications for angular sizes measured near that limit.
While our results show consistency between the direct measurements of effective temperature and indirectly determined literature values, a small difference exists between our radii measurements and theoretical estimates in the sense that for larger stars, the theoretical estimate falls slightly above the direct measurement.
From our effective temperature measurements, an empirical calibration of effective temperature versus $(V-K)_0$ color and spectral type is presented, with a spread of $\overline{\Delta T}_{\rm {(V-K)}_0} = 138$K over the range $(V-K)_0=0.0-4.0$ and $\overline{\Delta T}_{\rm {SpType}} = 105$K for F6V -- G5V.  No such calibration is possible for linear radius versus $(V-K)_0$ color, due to the large spread in radius values for any given $(V-K)_0$ color (presumably due to stellar evolution effects).  Among the stars considered, 55 Cnc is found to be problematic in terms of its interferometrically determined effective temperature, for reasons that are unclear.  Finally, the spectral energy distribution fitting tools employed in this investigation also enable indirect estimates of stellar angular size to be attempted for the full ensemble of stars known to host extrasolar planets, and this database of 166 stars is presented in the ``{\tt XO-Rad}'' appendix.



\acknowledgements

We would like to acknowledge constructive input and the occasional snide comment from David Ciardi.  This investigation has made extensive use of the {\tt sedFit} code, graciously provided by {\tt perl} guru Andrew F. Boden.
The preparation of this manuscript was greatly helped by the use of the
Extrasolar Planet Encyclopedia\footnote{Available at \url{http://exoplanet.eu}.}.
This research made use of the NASA/IPAC/NExScI Star and Exoplanet Database (NStED)\footnote{Available at \url{http://nsted.ipac.caltech.edu}.}, which is operated by the Jet Propulsion Laboratory, California Institute of Technology, under contract with the National Aeronautics and Space Administration.
This publication makes use of data products from the Two Micron All Sky
Survey, which is a joint project of the University of Massachusetts and the
Infrared Processing and Analysis Center/California Institute of Technology,
funded by the National Aeronautics and Space Administration and the National
Science Foundation.
The Palomar Testbed Interferometer is operated by the NASA Exoplanet Science Institute/Michelson Science Center on and the PTI collaboration and was constructed with funds from the Jet Propulsion Laboratory, Caltech as provided by the National Aeronautics and Space Administration.
This work has made use of services produced by the NASA Exoplanet Science Institute at the California Institute of Technology.


\begin{appendix} 

\section{The {\tt XO-Rad} Database}\label{sec_appendix}

For the full list of $\sim230$ stars found at the Extrasolar Planet Encyclopedia (as of 1 Feb 2008), we collected photometry and performed SED fits as described in the main manuscript in \S \ref{sec_sedfit}, and in detail in \citet{vvc07}.  64 of the stars have insufficient photometry and were dropped from the SED fitting.  The resultant 166 fits provide estimates of bolometric flux $F_{\rm BOL}$, $V$-band reddening $A_V$, angular size $\theta_{\rm EST}$, and linear radius $R_{\rm EST}$.  Effective temperatures are constrained to be those associated with the best fitting \citet{p98} empirical template.  Spectral types used in the SED fitting for EHS stars are those values found in the Exoplanet Encyclopedia, which is in turn based upon the respective source discovery papers cataloged therein.  The non-planet-hosting main sequence stars have their spectral types established from those values found in Hipparcos catalog \citep{plk97}. Linear radius is computed by combining the angular size estimates with the Hipparcos data found in \citet{vanLeeuwen2007A&A...474..653V}.  For a few of the stars, the linear radius is too large to be consistent with the main sequence spectral types indicated in the literature; for these objects, a second iteration on the SED fit is performed with a subgiant (luminosity class IV) template, resulting in a more appropriate set of fit parameters $\{ F_{\rm BOL}, A_V, \theta_{\rm EST}, R_{\rm EST}\}$.  The full {\tt XO-Rad} dataset of exoplanet radii is seen in Table \ref{table_XOrad}.

\end{appendix}

\bibliography{pti}

\clearpage
\begin{landscape}
\begin{deluxetable}{cccccccccccccc}
\tablecolumns{13}
\tablewidth{0pc}
\tablecaption{Observed and derived supporting parameters for luminosity class V stars.\label{tableData_LC5stars}}
\tablehead{
\colhead{Star ID} &
\colhead{$N(V^2)$} &
\colhead{$\theta_{\rm UD}$} &
\colhead{$\chi^2_\nu$} &
\colhead{${\overline \delta V^2}$} &
\colhead{$\theta_{\rm LD} / \theta_{\rm UD}$} &
\colhead{$\theta_{\rm LD}$} &
\colhead{$A_V$} &
\colhead{$F_{\rm BOL}$} &
\colhead{Spectral} &
\colhead{$V$} &
\colhead{$K$} \\
 &
\colhead{Points} &
\colhead{(mas)} &
&
&
&
\colhead{(mas)} &
\colhead{(mag)} &
\colhead{10$^{-8}$ erg cm$^{-2}$ s$^{-1}$} &
\colhead{Type} &
\colhead{(mag)} &
\colhead{(mag)} &
}
\tabletypesize{\scriptsize}
\startdata
\hline
\multicolumn{5}{l}{Control Sample: Stars not known to host planets:}\\
\hline
HD1326 & 216 & $1.009 \pm 0.009$ & $1.02$ & $0.058$ & $1.017$ & $1.027 \pm 0.059$ & $0.105 \pm 0.019$ & $5.79 \pm 0.13$ & - &  $8.15 \pm 0.05$ & $3.96 \pm 0.05$ \\
HD4628 & 98 & $0.911 \pm 0.013$ & $1.18$ & $0.042$ & $1.024$ & $0.933 \pm 0.064$ & $0.000 \pm 0.015$ & $17.12 \pm 0.29$ & K1V &  $5.74 \pm 0.05$ & $3.61 \pm 0.05$ \\
HD16160 & 42 & $0.820 \pm 0.045$ & $0.44$ & $0.055$ & $1.022$ & $0.838 \pm 0.069$ & $0.065 \pm 0.014$ & $17.93 \pm 0.31$ & K3V &  $5.83 \pm 0.05$ & $3.52 \pm 0.05$ \\
HD16895 & 118 & $1.067 \pm 0.016$ & $0.62$ & $0.036$ & $1.018$ & $1.086 \pm 0.056$ & $0.000 \pm 0.015$ & $58.06 \pm 0.99$ & F8 &  $4.11 \pm 0.05$ & $2.78 \pm 0.09$ \\
HD19373 & 14 & $1.304 \pm 0.022$ & $2.36$ & $0.052$ & $1.021$ & $1.331 \pm 0.050$ & $0.015 \pm 0.014$ & $63.24 \pm 0.95$ & F9.5V &  $4.05 \pm 0.05$ & $2.64 \pm 0.07$ \\
HD20630 & 2 & $0.878 \pm 0.068$ & $0.00$ & $0.000$ & $1.019$ & $0.895 \pm 0.070$ & $0.000 \pm 0.015$ & $32.46 \pm 0.55$ & G5V &  $4.85 \pm 0.05$ & $3.34 \pm 0.05$ \\
HD22484 & 8 & $0.897 \pm 0.122$ & $0.74$ & $0.064$ & $1.016$ & $0.911 \pm 0.123$ & $0.055 \pm 0.012$ & $52.99 \pm 0.77$ & F8V &  $4.30 \pm 0.05$ & $2.89 \pm 0.10$ \\
HD30652 & 38 & $1.388 \pm 0.024$ & $0.30$ & $0.053$ & $1.015$ & $1.409 \pm 0.048$ & $0.225 \pm 0.010$ & $164.90 \pm 2.51$ & F6V &  $3.18 \pm 0.05$ & $2.05 \pm 0.06$ \\
HD39587 & 84 & $1.102 \pm 0.018$ & $1.26$ & $0.068$ & $1.019$ & $1.124 \pm 0.056$ & $0.011 \pm 0.014$ & $46.45 \pm 0.76$ & G0IV-V &  $4.40 \pm 0.05$ & $2.99 \pm 0.07$ \\
HD87901 & 262 & $1.192 \pm 0.008$ & $1.18$ & $0.049$ & $1.015$ & $1.209 \pm 0.053$ & $0.150 \pm 0.010$ & $1997.00 \pm 26.62$ & B8IVn &  $1.40 \pm 0.05$ & $1.62 \pm 0.06$ \\
HD88230 & 64 & $1.208 \pm 0.013$ & $2.02$ & $0.096$ & $1.025$ & $1.238 \pm 0.053$ & $0.000 \pm 0.011$ & $15.23 \pm 0.06$ & K8V &  $6.61 \pm 0.05$ & $3.26 \pm 0.05$ \\
HD95735 & 80 & $1.417 \pm 0.009$ & $0.00$ & $0.001$ & $1.015$ & $1.439 \pm 0.048$ & $0.151 \pm 0.011$ & $11.49 \pm 0.05$ & Mb &  $7.51 \pm 0.05$ & $3.34 \pm 0.05$ \\
HD97603 & 126 & $1.180 \pm 0.010$ & $0.96$ & $0.046$ & $1.015$ & $1.198 \pm 0.053$ & $0.205 \pm 0.014$ & $299.60 \pm 5.69$ & A5 IV(n) &  $2.53 \pm 0.05$ & $2.24 \pm 0.06$ \\
HD102647 & 66 & $1.368 \pm 0.010$ & $0.51$ & $0.016$ & $1.015$ & $1.388 \pm 0.049$ & $0.038 \pm 0.015$ & $377.50 \pm 6.66$ & A3Va &  $2.13 \pm 0.05$ & $1.90 \pm 0.05$ \\
HD109358 & 166 & $1.117 \pm 0.008$ & $0.66$ & $0.036$ & $1.019$ & $1.138 \pm 0.055$ & $0.000 \pm 0.015$ & $52.12 \pm 0.87$ & G0V &  $4.25 \pm 0.05$ & $2.72 \pm 0.07$ \\
HD114710 & 28 & $1.052 \pm 0.014$ & $0.61$ & $0.037$ & $1.018$ & $1.071 \pm 0.057$ & $0.073 \pm 0.010$ & $55.28 \pm 0.64$ & G0 &  $4.25 \pm 0.05$ & $2.87 \pm 0.10$ \\
HD119850 & 142 & $0.811 \pm 0.011$ & $0.99$ & $0.062$ & $1.015$ & $0.823 \pm 0.069$ & $0.000 \pm 0.014$ & $4.06 \pm 0.03$ & K2 &  $8.50 \pm 0.05$ & $4.44 \pm 0.05$ \\
HD126660 & 134 & $1.111 \pm 0.014$ & $1.35$ & $0.049$ & $1.017$ & $1.130 \pm 0.055$ & $0.109 \pm 0.022$ & $69.46 \pm 1.99$ & F8 &  $4.05 \pm 0.05$ & $2.78 \pm 0.07$ \\
HD141004 & 6 & $0.824 \pm 0.118$ & $0.63$ & $0.024$ & $1.016$ & $0.838 \pm 0.120$ & $0.044 \pm 0.010$ & $46.01 \pm 0.58$ & G0IV-V &  $4.42 \pm 0.05$ & $2.96 \pm 0.08$ \\
HD142860 & 58 & $1.142 \pm 0.009$ & $0.42$ & $0.035$ & $1.017$ & $1.161 \pm 0.054$ & $0.053 \pm 0.014$ & $79.92 \pm 1.37$ & F5 &  $3.84 \pm 0.05$ & $2.62 \pm 0.06$ \\
HD149661 & 18 & $0.868 \pm 0.027$ & $1.99$ & $0.095$ & $1.023$ & $0.888 \pm 0.066$ & $0.324 \pm 0.015$ & $19.12 \pm 0.50$ & K1V &  $5.77 \pm 0.05$ & $3.83 \pm 0.05$ \\
HD157881 & 26 & $0.664 \pm 0.036$ & $0.35$ & $0.024$ & $1.023$ & $0.679 \pm 0.082$ & $0.000 \pm 0.014$ & $4.05 \pm 0.06$ & M1V &  $7.56 \pm 0.05$ & $4.14 \pm 0.05$ \\
HD185144 & 6 & $1.070 \pm 0.056$ & $0.59$ & $0.018$ & $1.021$ & $1.092 \pm 0.057$ & $0.000 \pm 0.013$ & $39.86 \pm 0.60$ & G9V &  $4.68 \pm 0.05$ & $2.83 \pm 0.08$ \\
HD201091 & 50 & $1.588 \pm 0.008$ & $0.47$ & $0.037$ & $1.025$ & $1.628 \pm 0.046$ & $0.000 \pm 0.011$ & $37.01 \pm 0.48$ & K5V &  $5.23 \pm 0.05$ & $2.68 \pm 0.05$ \\
HD201092 & 16 & $1.629 \pm 0.033$ & $1.05$ & $0.062$ & $1.023$ & $1.666 \pm 0.046$ & $0.232 \pm 0.012$ & $25.55 \pm 0.47$ & K7V &  $5.96 \pm 0.05$ & $2.32 \pm 0.05$ \\
HD210027 & 172 & $1.186 \pm 0.006$ & $2.44$ & $0.055$ & $1.017$ & $1.206 \pm 0.053$ & $0.000 \pm 0.011$ & $79.17 \pm 1.01$ & F5 &  $3.77 \pm 0.05$ & $2.50 \pm 0.07$ \\
HD215648 & 248 & $1.005 \pm 0.006$ & $1.09$ & $0.048$ & $1.017$ & $1.022 \pm 0.059$ & $0.101 \pm 0.017$ & $60.61 \pm 1.35$ & F5 &  $4.20 \pm 0.05$ & $2.87 \pm 0.08$ \\
HD222368 & 128 & $1.046 \pm 0.015$ & $0.91$ & $0.065$ & $1.016$ & $1.062 \pm 0.057$ & $0.148 \pm 0.014$ & $67.94 \pm 1.38$ & F8 &  $4.13 \pm 0.05$ & $2.75 \pm 0.08$ \\
\hline
\multicolumn{5}{l}{EHSA Sample: Known planet hosting stars (PTI):}\\
\hline
HD3651 & 222 & $0.670 \pm 0.080$ & $1.03$ & $0.082$ & $1.022$ & $0.685 \pm 0.081$ & $0.075 \pm 0.015$ & $13.84 \pm 0.23$ & K0V &  $5.88 \pm 0.05$ & $3.97 \pm 0.05$ \\
HD9826 & 540 & $1.004 \pm 0.058$ & $0.89$ & $0.035$ & $1.017$ & $1.021 \pm 0.059$ & $0.000 \pm 0.013$ & $60.68 \pm 0.91$ & G0 &  $4.10 \pm 0.05$ & $2.86 \pm 0.08$ \\
HD28305 & 32 & $2.422 \pm 0.044$ & $1.30$ & $0.028$ & $1.024$ & $2.481 \pm 0.045$ & $0.056 \pm 0.014$ & $127.10 \pm 2.03$ & K0III &  $3.53 \pm 0.05$ & $1.31 \pm 0.05$ \\
HD75732 & 16 & $0.796 \pm 0.069$ & $0.23$ & $0.018$ & $1.024$ & $0.816 \pm 0.071$ & $0.000 \pm 0.018$ & $13.32 \pm 0.26$ & K0IV-V &  $5.95 \pm 0.05$ & $4.01 \pm 0.22$ \\
HD95128 & 48 & $0.760 \pm 0.072$ & $1.34$ & $0.086$ & $1.018$ & $0.774 \pm 0.073$ & $0.123 \pm 0.024$ & $28.33 \pm 0.92$ & G0 &  $5.04 \pm 0.05$ & $3.75 \pm 0.34$ \\
HD117176 & 192 & $0.934 \pm 0.061$ & $1.40$ & $0.058$ & $1.021$ & $0.953 \pm 0.062$ & $0.121 \pm 0.015$ & $31.64 \pm 0.62$ & G0 &  $4.97 \pm 0.05$ & $3.24 \pm 0.05$ \\
HD120136 & 264 & $0.850 \pm 0.065$ & $0.98$ & $0.048$ & $1.016$ & $0.864 \pm 0.066$ & $0.219 \pm 0.018$ & $49.49 \pm 1.33$ & F5 &  $4.49 \pm 0.05$ & $3.36 \pm 0.05$ \\
HD143761 & 354 & $0.683 \pm 0.078$ & $0.31$ & $0.029$ & $1.019$ & $0.697 \pm 0.079$ & $0.096 \pm 0.016$ & $20.05 \pm 0.41$ & G0V &  $5.41 \pm 0.05$ & $3.89 \pm 0.05$ \\
HD217014 & 454 & $0.677 \pm 0.079$ & $1.28$ & $0.069$ & $1.019$ & $0.690 \pm 0.080$ & $0.043 \pm 0.009$ & $17.94 \pm 0.18$ & G3V &  $5.46 \pm 0.05$ & $3.99 \pm 0.05$ \\
\hline
\multicolumn{5}{l}{EHSA Sample: Known planet hosting stars (CHARA):}\\
\hline
HD3651 & $-$ & $0.773 \pm 0.026$ & $-$ & $-$ & $1.022$ & $0.790 \pm 0.027$ & $0.000 \pm 0.012$ & $13.64 \pm 0.18$ & K0V &  $5.88 \pm 0.05$ & $3.97 \pm 0.05$ \\
HD11964 & $-$ & $0.597 \pm 0.078$ & $-$ & $-$ & $1.023$ & $0.611 \pm 0.081$ & $0.437 \pm 0.010$ & $10.67 \pm 0.14$ & G5 &  $6.42 \pm 0.05$ & $4.49 \pm 0.02$ \\
HD19994 & $-$ & $0.774 \pm 0.026$ & $-$ & $-$ & $1.018$ & $0.788 \pm 0.026$ & $0.160 \pm 0.027$ & $28.79 \pm 0.83$ & F8V &  $5.08 \pm 0.05$ & $3.75 \pm 0.24$ \\
HD75732 & $-$ & $0.834 \pm 0.024$ & $-$ & $-$ & $1.024$ & $0.854 \pm 0.024$ & $0.111 \pm 0.022$ & $13.28 \pm 0.34$ & G8V &  $5.95 \pm 0.05$ & $4.02 \pm 0.04$ \\
HD143761 & $-$ & $0.673 \pm 0.043$ & $-$ & $-$ & $1.019$ & $0.686 \pm 0.044$ & $0.096 \pm 0.016$ & $20.05 \pm 0.41$ & G0V &  $5.41 \pm 0.05$ & $3.89 \pm 0.05$ \\
HD189733 & $-$ & $0.366 \pm 0.024$ & $-$ & $-$ & $1.030$ & $0.377 \pm 0.024$ & $0.101 \pm 0.018$ & $2.82 \pm 0.04$ & K1V &  $7.68 \pm 0.05$ & $5.54 \pm 0.02$ \\
HD217014 & $-$ & $0.733 \pm 0.026$ & $-$ & $-$ & $1.020$ & $0.748 \pm 0.027$ & $0.043 \pm 0.009$ & $17.94 \pm 0.18$ & G3V &  $5.46 \pm 0.05$ & $3.99 \pm 0.05$
\enddata
\end{deluxetable}
\clearpage
\end{landscape}

\LongTables
\begin{deluxetable}{rccrrrrrrr}
\tablecolumns{9}
\tablewidth{0pc}
\tablecaption{The {\tt XO-Rad} database: estimates of planetary host star bolometric flux, reddening, angular diameter, and linear radii from spectral energy distribution fitting.\label{table_XOrad}}
\tablehead{
\colhead{HD} &
\colhead{Template} &
\colhead{Template} &
\colhead{$\chi^2_\nu$} &
\colhead{$N_{\rm PHOT}$} &
\colhead{$F_{\rm BOL}$} &
\colhead{$A_{\rm V}$} &
\colhead{$\theta_{\rm EST}$} &
\colhead{$R_{\rm EST}$}\\
\colhead{Number} &
\colhead{} &
\colhead{$T_{\rm EFF}$ (K)} &
\colhead{} &
\colhead{} &
\colhead{($10^{-8}$ cm$^{-2}$ s$^{-1}$)} &
\colhead{(mag)} &
\colhead{(mas)} &
\colhead{($R_\odot$)}
}
\tabletypesize{\scriptsize}
\startdata
142 & F6V & $6280 \pm 70$ & 0.38 & 78 & $14.71 \pm 0.25$ & $0.09 \pm 0.01 $ & $0.533 \pm 0.013$ & $1.47 \pm 0.04$ \\
1237 & G8V & $5585 \pm 50$ & 3.24 & 23 & $7.01 \pm 0.09$ & $0.16 \pm 0.02 $ & $0.465 \pm 0.009$ & $0.88 \pm 0.02$ \\
2039 & G0IV & $5929 \pm 90$ & 1.08 & 23 & $0.69 \pm 0.01$ & $0.08 \pm 0.02 $ & $0.130 \pm 0.004$ & $1.43 \pm 0.15$ \\
2638 & G8V & $5585 \pm 50$ & 1.91 & 8 & $0.67 \pm 0.01$ & $0.38 \pm 0.02 $ & $0.158 \pm 0.003$ & $0.85 \pm 0.07$ \\
3651 & K0V & $5188 \pm 50$ & 2.52 & 135 & $14.81 \pm 0.23$ & $0.17 \pm 0.01 $ & $0.783 \pm 0.016$ & $0.93 \pm 0.02$ \\
4113 & G5V & $5585 \pm 50$ & 0.55 & 22 & $2.01 \pm 0.02$ & $0.07 \pm 0.01 $ & $0.249 \pm 0.005$ & $1.18 \pm 0.05$ \\
4208 & G5V & $5585 \pm 50$ & 1.44 & 46 & $2.20 \pm 0.03$ & $0.01 \pm 0.01 $ & $0.261 \pm 0.005$ & $0.91 \pm 0.03$ \\
4308 & G5V & $5636 \pm 50$ & 1.12 & 66 & $7.58 \pm 0.05$ & $0.18 \pm 0.01 $ & $0.447 \pm 0.008$ & $1.06 \pm 0.02$ \\
5319 & G5IV & $5598 \pm 80$ & 5.23 & 64 & $3.59 \pm 0.09$ & $0.92 \pm 0.01 $ & $0.331 \pm 0.010$ & $4.08 \pm 0.42$ \\
6434 & G2V & $5807 \pm 50$ & 1.65 & 41 & $2.43 \pm 0.03$ & $0.09 \pm 0.01 $ & $0.253 \pm 0.005$ & $1.13 \pm 0.03$ \\
8574 & F8V & $6040 \pm 50$ & 0.06 & 23 & $3.92 \pm 0.03$ & $0.05 \pm 0.01 $ & $0.297 \pm 0.005$ & $1.42 \pm 0.04$ \\
9826 & F8IV & $6152 \pm 100$ & 0.64 & 134 & $60.60 \pm 0.89$ & $0.00 \pm 0.01 $ & $1.130 \pm 0.038$ & $1.64 \pm 0.05$ \\
10647 & F8V & $6040 \pm 50$ & 0.57 & 61 & $16.41 \pm 0.60$ & $0.00 \pm 0.03 $ & $0.608 \pm 0.015$ & $1.14 \pm 0.03$ \\
10697 & G5IV & $5689 \pm 85$ & 0.28 & 47 & $9.48 \pm 0.13$ & $0.15 \pm 0.01 $ & $0.521 \pm 0.016$ & $1.83 \pm 0.06$ \\
11506 & G0V & $5807 \pm 50$ & 0.66 & 30 & $2.56 \pm 0.02$ & $0.01 \pm 0.01 $ & $0.260 \pm 0.005$ & $1.45 \pm 0.05$ \\
11964 & G5IV & $5598 \pm 80$ & 0.47 & 52 & $9.80 \pm 0.11$ & $0.30 \pm 0.01 $ & $0.547 \pm 0.016$ & $1.93 \pm 0.07$ \\
11977 & G8III & $5012 \pm 150$ & 0.80 & 36 & $46.52 \pm 1.63$ & $0.08 \pm 0.03 $ & $1.490 \pm 0.093$ & $10.75 \pm 0.68$ \\
12661 & K0V & $5333 \pm 50$ & 1.11 & 26 & $3.08 \pm 0.03$ & $0.10 \pm 0.01 $ & $0.308 \pm 0.006$ & $1.16 \pm 0.03$ \\
13189 & K3III & $4365 \pm 100$ & 0.98 & 7 & $6.85 \pm 0.27$ & $0.60 \pm 0.04 $ & $0.753 \pm 0.038$ & $45.53 \pm 18.81$ \\
13445 & K0V & $5188 \pm 50$ & 0.52 & 82 & $11.62 \pm 0.28$ & $0.10 \pm 0.02 $ & $0.694 \pm 0.016$ & $0.81 \pm 0.02$ \\
16141 & G5IV & $5689 \pm 85$ & 0.53 & 59 & $5.06 \pm 0.05$ & $0.00 \pm 0.01 $ & $0.381 \pm 0.012$ & $1.60 \pm 0.06$ \\
16175 & F8IV & $6152 \pm 100$ & 0.94 & 15 & $3.54 \pm 0.05$ & $0.18 \pm 0.02 $ & $0.272 \pm 0.009$ & $1.69 \pm 0.09$ \\
17051 & F8V & $6040 \pm 50$ & 0.24 & 73 & $18.57 \pm 0.40$ & $0.03 \pm 0.02 $ & $0.647 \pm 0.013$ & $1.20 \pm 0.02$ \\
17092 & K0III & $4853 \pm 130$ & 8.12 & 5 & $5.21 \pm 0.05$ & $0.80 \pm 0.04 $ & $0.531 \pm 0.029$ & $6.22 \pm 3.74$ \\
17156 & F8IV & $6152 \pm 100$ & 3.13 & 5 & $1.48 \pm 0.02$ & $0.10 \pm 0.04 $ & $0.176 \pm 0.006$ & $1.42 \pm 0.09$ \\
19994 & F8IV & $6152 \pm 100$ & 0.48 & 82 & $26.61 \pm 0.95$ & $0.09 \pm 0.03 $ & $0.747 \pm 0.028$ & $1.82 \pm 0.07$ \\
20367 & G0V & $5807 \pm 50$ & 1.12 & 37 & $7.35 \pm 0.07$ & $0.01 \pm 0.01 $ & $0.441 \pm 0.008$ & $1.27 \pm 0.03$ \\
20782 & G0V & $5807 \pm 50$ & 1.58 & 34 & $3.56 \pm 0.02$ & $0.20 \pm 0.01 $ & $0.306 \pm 0.005$ & $1.17 \pm 0.03$ \\
22049 & K2V & $4887 \pm 50$ & 2.15 & 201 & $108.00 \pm 1.06$ & $0.00 \pm 0.01 $ & $2.380 \pm 0.050$ & $0.82 \pm 0.02$ \\
23079 & F8V & $6040 \pm 50$ & 1.02 & 26 & $4.03 \pm 0.03$ & $0.07 \pm 0.01 $ & $0.301 \pm 0.005$ & $1.10 \pm 0.02$ \\
23127 & G2IV & $5689 \pm 85$ & 1.70 & 30 & $0.97 \pm 0.01$ & $0.02 \pm 0.01 $ & $0.167 \pm 0.005$ & $1.77 \pm 0.13$ \\
27442 & K2III & $4457 \pm 110$ & 1.80 & 66 & $67.54 \pm 1.33$ & $0.02 \pm 0.02 $ & $2.270 \pm 0.114$ & $4.46 \pm 0.22$ \\
27894 & K2V & $4887 \pm 50$ & 1.34 & 10 & $0.64 \pm 0.01$ & $0.09 \pm 0.02 $ & $0.183 \pm 0.004$ & $0.86 \pm 0.04$ \\
28305 & G8III & $5012 \pm 150$ & 3.62 & 85 & $136.70 \pm 2.36$ & $0.21 \pm 0.01 $ & $2.550 \pm 0.154$ & $12.34 \pm 0.76$ \\
30177 & G8V & $5585 \pm 50$ & 1.03 & 16 & $1.37 \pm 0.03$ & $0.22 \pm 0.02 $ & $0.205 \pm 0.004$ & $1.17 \pm 0.05$ \\
33283 & G0IV & $5929 \pm 90$ & 1.09 & 29 & $1.52 \pm 0.02$ & $0.00 \pm 0.02 $ & $0.192 \pm 0.006$ & $1.95 \pm 0.13$ \\
33564 & F6V & $6531 \pm 70$ & 1.01 & 34 & $25.39 \pm 0.33$ & $0.07 \pm 0.01 $ & $0.647 \pm 0.015$ & $1.45 \pm 0.03$ \\
37124 & G2V & $5636 \pm 50$ & 0.65 & 46 & $2.67 \pm 0.02$ & $0.11 \pm 0.01 $ & $0.282 \pm 0.005$ & $1.02 \pm 0.03$ \\
37605 & K0V & $5188 \pm 50$ & 3.21 & 31 & $0.99 \pm 0.01$ & $0.05 \pm 0.01 $ & $0.203 \pm 0.004$ & $0.96 \pm 0.05$ \\
38529 & G2IV & $5689 \pm 85$ & 0.55 & 32 & $13.92 \pm 0.49$ & $0.25 \pm 0.03 $ & $0.632 \pm 0.022$ & $2.67 \pm 0.10$ \\
39091 & G0V & $5807 \pm 50$ & 0.22 & 68 & $15.35 \pm 0.31$ & $0.06 \pm 0.02 $ & $0.636 \pm 0.013$ & $1.25 \pm 0.03$ \\
40979 & F8V & $6040 \pm 50$ & 0.26 & 30 & $5.27 \pm 0.05$ & $0.01 \pm 0.01 $ & $0.345 \pm 0.006$ & $1.23 \pm 0.03$ \\
41004 & K0V & $5188 \pm 50$ & 1.08 & 28 & $1.43 \pm 0.03$ & $0.38 \pm 0.02 $ & $0.243 \pm 0.005$ & $1.07 \pm 0.04$ \\
41004 & K0V & $5188 \pm 50$ & 1.08 & 28 & $1.43 \pm 0.03$ & $0.38 \pm 0.02 $ & $0.243 \pm 0.005$ & $1.07 \pm 0.04$ \\
43691 & F8IV & $6152 \pm 100$ & 0.41 & 15 & $1.60 \pm 0.02$ & $0.03 \pm 0.02 $ & $0.183 \pm 0.006$ & $1.58 \pm 0.12$ \\
44627 & K2V & $5188 \pm 50$ & 0.10 & 16 & $0.83 \pm 0.01$ & $0.27 \pm 0.02 $ & $0.185 \pm 0.004$ & $0.92 \pm 0.03$ \\
45350 & G5V & $5585 \pm 50$ & 0.64 & 21 & $2.18 \pm 0.01$ & $0.16 \pm 0.01 $ & $0.259 \pm 0.005$ & $1.36 \pm 0.05$ \\
46375 & G8IV & $5012 \pm 85$ & 3.89 & 16 & $2.13 \pm 0.03$ & $0.06 \pm 0.02 $ & $0.284 \pm 0.008$ & $1.06 \pm 0.05$ \\
47536 & K0III & $4853 \pm 130$ & 4.79 & 42 & $48.02 \pm 3.10$ & $0.72 \pm 0.03 $ & $1.610 \pm 0.101$ & $21.36 \pm 1.47$ \\
49674 & G0V & $5807 \pm 50$ & 1.50 & 15 & $2.14 \pm 0.02$ & $0.44 \pm 0.02 $ & $0.238 \pm 0.004$ & $1.13 \pm 0.05$ \\
50499 & F8IV & $6152 \pm 100$ & 0.75 & 23 & $3.57 \pm 0.02$ & $0.10 \pm 0.01 $ & $0.273 \pm 0.009$ & $1.36 \pm 0.05$ \\
50554 & F8V & $6040 \pm 50$ & 0.40 & 32 & $5.17 \pm 0.04$ & $0.08 \pm 0.01 $ & $0.342 \pm 0.006$ & $1.10 \pm 0.03$ \\
52265 & G0IV & $6152 \pm 100$ & 0.41 & 88 & $8.00 \pm 0.07$ & $0.03 \pm 0.01 $ & $0.410 \pm 0.013$ & $1.28 \pm 0.04$ \\
59686 & K2III & $4656 \pm 120$ & 5.41 & 17 & $23.68 \pm 1.17$ & $0.08 \pm 0.05 $ & $1.230 \pm 0.070$ & $12.83 \pm 0.81$ \\
61098 & B6IV & $12589 \pm 300$ & 3.33 & 10 & $3.27 \pm 0.05$ & $0.94 \pm 0.03 $ & $0.063 \pm 0.003$ & $1.10 \pm 0.49$ \\
62509 & K0III & $4853 \pm 130$ & 1.59 & 101 & $1,234.00 \pm 22.35$ & $0.10 \pm 0.02 $ & $8.170 \pm 0.444$ & $9.11 \pm 0.50$ \\
65216 & G5V & $5636 \pm 50$ & 0.36 & 31 & $1.81 \pm 0.03$ & $0.03 \pm 0.02 $ & $0.232 \pm 0.004$ & $0.89 \pm 0.03$ \\
66428 & G5V & $5585 \pm 50$ & 3.51 & 33 & $1.40 \pm 0.02$ & $0.03 \pm 0.01 $ & $0.208 \pm 0.004$ & $1.23 \pm 0.08$ \\
68988 & G0V & $5807 \pm 50$ & 0.67 & 15 & $1.57 \pm 0.02$ & $0.18 \pm 0.02 $ & $0.203 \pm 0.004$ & $1.19 \pm 0.05$ \\
69830 & G8V & $5333 \pm 50$ & 0.36 & 104 & $12.20 \pm 0.25$ & $0.03 \pm 0.02 $ & $0.673 \pm 0.014$ & $0.90 \pm 0.02$ \\
70573 & G0V & $5807 \pm 50$ & 0.42 & 23 & $1.00 \pm 0.01$ & $0.15 \pm 0.02 $ & $0.163 \pm 0.003$ & $1.55 \pm 1.47$ \\
70642 & G5V & $5585 \pm 50$ & 0.40 & 40 & $3.80 \pm 0.03$ & $0.04 \pm 0.01 $ & $0.342 \pm 0.006$ & $1.03 \pm 0.02$ \\
72659 & F8IV & $6152 \pm 100$ & 0.79 & 23 & $2.94 \pm 0.03$ & $0.11 \pm 0.01 $ & $0.248 \pm 0.008$ & $1.33 \pm 0.07$ \\
73108 & K1III & $4656 \pm 120$ & 1.02 & 54 & $96.13 \pm 5.27$ & $0.69 \pm 0.02 $ & $2.480 \pm 0.145$ & $20.95 \pm 1.30$ \\
73256 & G8IV & $5598 \pm 80$ & 0.13 & 23 & $1.92 \pm 0.02$ & $0.17 \pm 0.02 $ & $0.242 \pm 0.007$ & $0.98 \pm 0.04$ \\
73526 & G5IV & $5598 \pm 80$ & 0.40 & 26 & $0.72 \pm 0.01$ & $0.00 \pm 0.02 $ & $0.148 \pm 0.004$ & $1.60 \pm 0.17$ \\
74156 & F8IV & $6152 \pm 100$ & 0.52 & 30 & $2.44 \pm 0.03$ & $0.06 \pm 0.01 $ & $0.226 \pm 0.007$ & $1.57 \pm 0.08$ \\
75289 & F8V & $6040 \pm 50$ & 1.28 & 45 & $7.57 \pm 0.06$ & $0.03 \pm 0.01 $ & $0.413 \pm 0.007$ & $1.30 \pm 0.03$ \\
75732 & G8V & $5333 \pm 50$ & 1.03 & 49 & $14.25 \pm 0.41$ & $0.23 \pm 0.02 $ & $0.727 \pm 0.017$ & $0.97 \pm 0.02$ \\
75898 & F8IV & $6152 \pm 100$ & 0.72 & 15 & $1.70 \pm 0.02$ & $0.11 \pm 0.02 $ & $0.189 \pm 0.006$ & $1.54 \pm 0.11$ \\
76700 & G5IV & $5598 \pm 80$ & 1.03 & 32 & $1.54 \pm 0.02$ & $0.00 \pm 0.01 $ & $0.217 \pm 0.006$ & $1.41 \pm 0.06$ \\
80606 & G5V & $5585 \pm 50$ & 1.18 & 15 & $0.80 \pm 0.01$ & $0.19 \pm 0.02 $ & $0.157 \pm 0.003$ & $0.99 \pm 0.33$ \\
81040 & G0V & $5807 \pm 50$ & 0.34 & 26 & $2.67 \pm 0.03$ & $0.23 \pm 0.01 $ & $0.265 \pm 0.005$ & $0.94 \pm 0.04$ \\
82943 & F8V & $6040 \pm 50$ & 7.63 & 17 & $6.79 \pm 0.05$ & $0.13 \pm 0.02 $ & $0.391 \pm 0.007$ & $1.16 \pm 0.02$ \\
86081 & F8IV & $6152 \pm 100$ & 2.64 & 5 & $0.91 \pm 0.02$ & $0.17 \pm 0.05 $ & $0.138 \pm 0.005$ & $1.42 \pm 0.14$ \\
88133 & G5IV & $5598 \pm 80$ & 1.74 & 28 & $2.13 \pm 0.01$ & $0.28 \pm 0.01 $ & $0.255 \pm 0.007$ & $2.23 \pm 0.17$ \\
89307 & G0V & $5807 \pm 50$ & 0.65 & 35 & $4.46 \pm 0.04$ & $0.06 \pm 0.01 $ & $0.343 \pm 0.006$ & $1.19 \pm 0.03$ \\
89744 & F5IV & $6562 \pm 150$ & 0.46 & 36 & $16.92 \pm 0.31$ & $0.33 \pm 0.01 $ & $0.523 \pm 0.024$ & $2.22 \pm 0.11$ \\
93083 & K2V & $5188 \pm 50$ & 1.88 & 18 & $1.85 \pm 0.02$ & $0.33 \pm 0.02 $ & $0.325 \pm 0.016$ & $0.97 \pm 0.05$ \\
93989 & B9III & $11092 \pm 1000$ & 2.60 & 14 & $2.02 \pm 0.01$ & $0.48 \pm 0.02 $ & $0.063 \pm 0.011$ & $16.59 \pm 29.29$ \\
94346 & B5III & $14791 \pm 1200$ & 1.30 & 21 & $12.69 \pm 0.12$ & $0.70 \pm 0.02 $ & $0.089 \pm 0.015$ & $10.32 \pm 5.59$ \\
95128 & G0V & $5807 \pm 50$ & 0.35 & 109 & $28.33 \pm 0.91$ & $0.12 \pm 0.02 $ & $0.865 \pm 0.020$ & $1.31 \pm 0.03$ \\
99109 & K0V & $5188 \pm 50$ & 3.46 & 8 & $7.06 \pm 0.01$ & $0.06 \pm 0.02 $ & $0.171 \pm 0.004$ & $0.92 \pm 0.07$ \\
99492 & K2V & $4887 \pm 50$ & 1.27 & 44 & $3.41 \pm 0.08$ & $0.14 \pm 0.02 $ & $0.423 \pm 0.010$ & $0.82 \pm 0.03$ \\
100777 & G8V & $5333 \pm 50$ & 3.14 & 11 & $1.14 \pm 0.02$ & $0.05 \pm 0.04 $ & $0.206 \pm 0.004$ & $1.10 \pm 0.06$ \\
101930 & K2V & $5188 \pm 50$ & 3.22 & 17 & $1.90 \pm 0.02$ & $0.26 \pm 0.02 $ & $0.281 \pm 0.006$ & $0.88 \pm 0.03$ \\
102117 & G5V & $5585 \pm 50$ & 0.47 & 15 & $3.03 \pm 0.03$ & $0.11 \pm 0.02 $ & $0.306 \pm 0.006$ & $1.31 \pm 0.04$ \\
102195 & K0V & $5188 \pm 50$ & 1.89 & 13 & $1.74 \pm 0.02$ & $0.00 \pm 0.02 $ & $0.268 \pm 0.005$ & $0.85 \pm 0.03$ \\
104985 & G8III & $5012 \pm 150$ & 0.49 & 22 & $23.24 \pm 1.78$ & $0.45 \pm 0.05 $ & $1.050 \pm 0.075$ & $10.97 \pm 0.82$ \\
107148 & G5V & $5636 \pm 50$ & 1.39 & 24 & $1.63 \pm 0.02$ & $0.02 \pm 0.02 $ & $0.220 \pm 0.004$ & $1.21 \pm 0.05$ \\
108147 & F8V & $6280 \pm 70$ & 0.47 & 39 & $4.57 \pm 0.03$ & $0.12 \pm 0.01 $ & $0.297 \pm 0.007$ & $1.22 \pm 0.03$ \\
108874 & G5V & $5585 \pm 50$ & 1.63 & 11 & $0.97 \pm 0.01$ & $0.16 \pm 0.03 $ & $0.173 \pm 0.003$ & $1.17 \pm 0.08$ \\
109749 & G2V & $5807 \pm 50$ & 0.66 & 31 & $1.62 \pm 0.02$ & $0.20 \pm 0.01 $ & $0.207 \pm 0.004$ & $1.25 \pm 0.09$ \\
111232 & G8V & $5585 \pm 50$ & 1.67 & 28 & $2.75 \pm 0.03$ & $0.06 \pm 0.02 $ & $0.291 \pm 0.005$ & $0.92 \pm 0.02$ \\
114386 & K3V & $4498 \pm 50$ & 1.21 & 10 & $1.14 \pm 0.01$ & $0.00 \pm 0.02 $ & $0.289 \pm 0.007$ & $0.90 \pm 0.04$ \\
114729 & F8IV & $6152 \pm 100$ & 0.95 & 64 & $6.21 \pm 0.06$ & $0.13 \pm 0.01 $ & $0.361 \pm 0.012$ & $1.40 \pm 0.05$ \\
114762 & F8V & $6040 \pm 50$ & 1.88 & 100 & $3.64 \pm 0.03$ & $0.09 \pm 0.01 $ & $0.286 \pm 0.005$ & $1.19 \pm 0.04$ \\
114783 & K0V & $5188 \pm 50$ & 3.25 & 33 & $3.38 \pm 0.03$ & $0.28 \pm 0.01 $ & $0.374 \pm 0.007$ & $0.83 \pm 0.02$ \\
117176 & G2IV & $5689 \pm 85$ & 0.46 & 89 & $32.92 \pm 0.69$ & $0.18 \pm 0.02 $ & $0.971 \pm 0.031$ & $1.88 \pm 0.06$ \\
117207 & G5V & $5585 \pm 50$ & 0.67 & 22 & $3.55 \pm 0.03$ & $0.07 \pm 0.01 $ & $0.331 \pm 0.006$ & $1.18 \pm 0.03$ \\
117618 & G0V & $5807 \pm 50$ & 0.84 & 15 & $3.77 \pm 0.04$ & $0.06 \pm 0.02 $ & $0.316 \pm 0.006$ & $1.29 \pm 0.04$ \\
118203 & G2IV & $5689 \pm 85$ & 4.86 & 5 & $1.63 \pm 0.02$ & $0.00 \pm 0.04 $ & $0.216 \pm 0.007$ & $2.06 \pm 0.14$ \\
120136 & F5IV & $6562 \pm 150$ & 0.39 & 121 & $49.49 \pm 1.31$ & $0.22 \pm 0.02 $ & $0.895 \pm 0.043$ & $1.50 \pm 0.07$ \\
121504 & G2V & $5807 \pm 50$ & 0.59 & 31 & $2.72 \pm 0.03$ & $0.09 \pm 0.01 $ & $0.268 \pm 0.005$ & $1.30 \pm 0.05$ \\
125612 & G2V & $5807 \pm 50$ & 0.18 & 24 & $1.42 \pm 0.02$ & $0.16 \pm 0.02 $ & $0.194 \pm 0.004$ & $1.13 \pm 0.07$ \\
128311 & K0V & $5188 \pm 50$ & 4.89 & 34 & $4.71 \pm 0.04$ & $0.59 \pm 0.01 $ & $0.442 \pm 0.009$ & $0.78 \pm 0.02$ \\
130322 & G0III & $4853 \pm 130$ & 1.65 & 24 & $1.77 \pm 0.02$ & $0.07 \pm 0.02 $ & $0.232 \pm 0.015$ & $0.79 \pm 0.06$ \\
132406 & G0V & $5807 \pm 50$ & 1.11 & 9 & $1.39 \pm 0.01$ & $0.27 \pm 0.02 $ & $0.192 \pm 0.003$ & $1.40 \pm 0.06$ \\
134987 & G5V & $5585 \pm 50$ & 0.07 & 34 & $6.90 \pm 0.14$ & $0.00 \pm 0.02 $ & $0.445 \pm 0.014$ & $1.25 \pm 0.04$ \\
136118 & F8V & $6040 \pm 50$ & 1.13 & 23 & $4.56 \pm 0.05$ & $0.04 \pm 0.02 $ & $0.321 \pm 0.006$ & $1.61 \pm 0.05$ \\
141937 & G0V & $5807 \pm 50$ & 0.77 & 22 & $3.80 \pm 0.03$ & $0.16 \pm 0.01 $ & $0.317 \pm 0.006$ & $1.10 \pm 0.03$ \\
142415 & G0V & $5807 \pm 50$ & 0.30 & 38 & $3.34 \pm 0.03$ & $0.09 \pm 0.01 $ & $0.297 \pm 0.005$ & $1.09 \pm 0.03$ \\
143761 & G0V & $5807 \pm 50$ & 0.41 & 133 & $20.05 \pm 0.40$ & $0.10 \pm 0.02 $ & $0.728 \pm 0.015$ & $1.35 \pm 0.03$ \\
145675 & K0V & $5188 \pm 50$ & 2.65 & 46 & $6.58 \pm 0.03$ & $0.06 \pm 0.01 $ & $0.522 \pm 0.010$ & $0.99 \pm 0.02$ \\
147513 & G0V & $5807 \pm 50$ & 0.41 & 80 & $20.78 \pm 0.55$ & $0.12 \pm 0.02 $ & $0.741 \pm 0.016$ & $1.02 \pm 0.02$ \\
149026 & G0IV & $6152 \pm 100$ & 1.48 & 16 & $1.41 \pm 0.01$ & $0.05 \pm 0.02 $ & $0.172 \pm 0.006$ & $1.47 \pm 0.09$ \\
149143 & F8IV & $6152 \pm 100$ & 0.94 & 23 & $2.07 \pm 0.03$ & $0.21 \pm 0.02 $ & $0.208 \pm 0.007$ & $1.39 \pm 0.09$ \\
150706 & G0V & $5807 \pm 50$ & 0.77 & 34 & $4.54 \pm 0.02$ & $0.10 \pm 0.01 $ & $0.346 \pm 0.006$ & $1.05 \pm 0.02$ \\
154345 & G8V & $5585 \pm 50$ & 0.44 & 67 & $6.64 \pm 0.06$ & $0.20 \pm 0.01 $ & $0.452 \pm 0.008$ & $0.90 \pm 0.02$ \\
154857 & G2IV & $5689 \pm 85$ & 0.90 & 19 & $4.09 \pm 0.03$ & $0.14 \pm 0.02 $ & $0.342 \pm 0.010$ & $2.36 \pm 0.13$ \\
157931 & G8IV & $5309 \pm 75$ & 0.73 & 28 & $1.23 \pm 0.01$ & $0.08 \pm 0.01 $ & $0.216 \pm 0.006$ & $2.69 \pm 0.41$ \\
159868 & G2IV & $5689 \pm 85$ & 0.51 & 81 & $4.12 \pm 0.05$ & $0.19 \pm 0.01 $ & $0.343 \pm 0.011$ & $2.17 \pm 0.12$ \\
160691 & G2IV & $5689 \pm 85$ & 0.28 & 69 & $26.02 \pm 0.89$ & $0.10 \pm 0.03 $ & $0.863 \pm 0.030$ & $1.44 \pm 0.05$ \\
164922 & K0V & $5188 \pm 50$ & 1.30 & 59 & $4.58 \pm 0.03$ & $0.01 \pm 0.01 $ & $0.435 \pm 0.009$ & $1.04 \pm 0.02$ \\
167042 & K1III & $4853 \pm 130$ & 1.35 & 35 & $14.00 \pm 0.60$ & $0.00 \pm 0.04 $ & $0.870 \pm 0.050$ & $4.70 \pm 0.28$ \\
168443 & G2IV & $5689 \pm 85$ & 0.33 & 31 & $5.40 \pm 0.02$ & $0.18 \pm 0.01 $ & $0.393 \pm 0.012$ & $1.58 \pm 0.06$ \\
168746 & G5V & $5585 \pm 50$ & 0.68 & 23 & $2.05 \pm 0.03$ & $0.10 \pm 0.02 $ & $0.251 \pm 0.005$ & $1.15 \pm 0.04$ \\
169830 & F5IV & $6562 \pm 150$ & 0.67 & 34 & $12.90 \pm 0.09$ & $0.20 \pm 0.01 $ & $0.457 \pm 0.021$ & $1.80 \pm 0.09$ \\
171028 & G0V & $5807 \pm 50$ & 0.77 & 23 & $1.77 \pm 0.03$ & $0.31 \pm 0.02 $ & $0.216 \pm 0.004$ & $2.55 \pm 2.19$ \\
175541 & G5IV & $5598 \pm 80$ & 0.65 & 25 & $2.99 \pm 0.06$ & $0.61 \pm 0.01 $ & $0.302 \pm 0.009$ & $4.13 \pm 0.51$ \\
177830 & G8IV & $5309 \pm 75$ & 0.80 & 14 & $6.90 \pm 0.23$ & $0.66 \pm 0.03 $ & $0.511 \pm 0.017$ & $3.25 \pm 0.16$ \\
178911 & F8IV & $6152 \pm 100$ & 0.50 & 32 & $6.31 \pm 0.05$ & $0.23 \pm 0.01 $ & $0.364 \pm 0.012$ & $2.05 \pm 0.26$ \\
179949 & F8V & $6040 \pm 50$ & 0.78 & 50 & $8.33 \pm 0.08$ & $0.00 \pm 0.01 $ & $0.433 \pm 0.007$ & $1.28 \pm 0.03$ \\
183263 & G2IV & $5929 \pm 90$ & 1.17 & 24 & $1.85 \pm 0.02$ & $0.03 \pm 0.01 $ & $0.212 \pm 0.007$ & $1.26 \pm 0.08$ \\
185269 & G0IV & $6152 \pm 100$ & 0.49 & 30 & $6.15 \pm 0.06$ & $0.13 \pm 0.01 $ & $0.359 \pm 0.012$ & $1.94 \pm 0.08$ \\
186427 & G2V & $5636 \pm 50$ & 1.10 & 129 & $8.86 \pm 0.04$ & $0.02 \pm 0.01 $ & $0.513 \pm 0.009$ & $1.17 \pm 0.02$ \\
187085 & G0V & $6040 \pm 50$ & 0.51 & 16 & $3.48 \pm 0.04$ & $0.04 \pm 0.02 $ & $0.280 \pm 0.005$ & $1.33 \pm 0.05$ \\
187123 & G5V & $5636 \pm 50$ & 1.62 & 17 & $2.29 \pm 0.01$ & $0.19 \pm 0.02 $ & $0.246 \pm 0.004$ & $1.28 \pm 0.04$ \\
189733 & G5V & $5585 \pm 50$ & 2.45 & 16 & $4.36 \pm 0.10$ & $0.74 \pm 0.02 $ & $0.367 \pm 0.008$ & $0.77 \pm 0.02$ \\
190228 & G5IV & $5598 \pm 80$ & 1.60 & 21 & $4.51 \pm 0.03$ & $0.29 \pm 0.01 $ & $0.371 \pm 0.011$ & $2.46 \pm 0.12$ \\
190360 & G5V & $5585 \pm 50$ & 0.52 & 81 & $16.38 \pm 0.37$ & $0.19 \pm 0.02 $ & $0.711 \pm 0.015$ & $1.21 \pm 0.03$ \\
190647 & G5IV & $5598 \pm 80$ & 0.68 & 22 & $2.14 \pm 0.03$ & $0.00 \pm 0.02 $ & $0.256 \pm 0.007$ & $1.58 \pm 0.09$ \\
192263 & K2V & $4887 \pm 50$ & 1.18 & 30 & $2.57 \pm 0.02$ & $0.02 \pm 0.01 $ & $0.368 \pm 0.008$ & $0.76 \pm 0.02$ \\
192699 & G5IV & $5598 \pm 80$ & 0.21 & 21 & $11.42 \pm 0.38$ & $0.47 \pm 0.02 $ & $0.591 \pm 0.020$ & $4.17 \pm 0.21$ \\
195019 & G2V & $5636 \pm 50$ & 0.44 & 40 & $5.09 \pm 0.02$ & $0.05 \pm 0.01 $ & $0.389 \pm 0.007$ & $1.61 \pm 0.07$ \\
196050 & G2IV & $5689 \pm 85$ & 1.03 & 34 & $2.61 \pm 0.03$ & $0.00 \pm 0.01 $ & $0.273 \pm 0.008$ & $1.47 \pm 0.07$ \\
196885 & F8IV & $6562 \pm 150$ & 1.32 & 43 & $9.29 \pm 0.10$ & $0.31 \pm 0.01 $ & $0.388 \pm 0.018$ & $1.40 \pm 0.07$ \\
202206 & G5V & $5585 \pm 50$ & 1.23 & 30 & $1.66 \pm 0.01$ & $0.03 \pm 0.01 $ & $0.226 \pm 0.004$ & $1.10 \pm 0.05$ \\
208487 & G2V & $5807 \pm 50$ & 0.74 & 22 & $2.63 \pm 0.02$ & $0.00 \pm 0.02 $ & $0.244 \pm 0.004$ & $1.20 \pm 0.04$ \\
209458 & G0V & $6040 \pm 50$ & 0.14 & 24 & $2.37 \pm 0.03$ & $0.03 \pm 0.02 $ & $0.231 \pm 0.004$ & $1.23 \pm 0.05$ \\
210277 & G0V & $5807 \pm 50$ & 2.33 & 36 & $9.55 \pm 0.15$ & $0.55 \pm 0.01 $ & $0.502 \pm 0.010$ & $1.16 \pm 0.03$ \\
210702 & K1III & $4853 \pm 130$ & 1.12 & 63 & $14.27 \pm 0.43$ & $0.01 \pm 0.03 $ & $0.879 \pm 0.049$ & $5.20 \pm 0.31$ \\
212301 & F8V & $6280 \pm 70$ & 0.15 & 24 & $2.29 \pm 0.02$ & $0.15 \pm 0.01 $ & $0.210 \pm 0.005$ & $1.24 \pm 0.05$ \\
213240 & F8IV & $6152 \pm 100$ & 1.03 & 38 & $5.33 \pm 0.04$ & $0.11 \pm 0.01 $ & $0.334 \pm 0.011$ & $1.46 \pm 0.06$ \\
216435 & G0IV & $5929 \pm 90$ & 0.38 & 61 & $9.90 \pm 0.11$ & $0.00 \pm 0.01 $ & $0.490 \pm 0.015$ & $1.72 \pm 0.06$ \\
216437 & G0IV & $5929 \pm 90$ & 0.93 & 74 & $10.52 \pm 0.19$ & $0.10 \pm 0.02 $ & $0.506 \pm 0.016$ & $1.46 \pm 0.05$ \\
216770 & G8V & $5333 \pm 50$ & 2.18 & 32 & $1.69 \pm 0.01$ & $0.10 \pm 0.01 $ & $0.250 \pm 0.005$ & $0.96 \pm 0.03$ \\
217014 & G2V & $5636 \pm 50$ & 1.86 & 214 & $17.94 \pm 0.18$ & $0.04 \pm 0.01 $ & $0.731 \pm 0.014$ & $1.23 \pm 0.02$ \\
217107 & G8IV & $5598 \pm 80$ & 0.43 & 43 & $9.32 \pm 0.12$ & $0.01 \pm 0.01 $ & $0.534 \pm 0.016$ & $1.14 \pm 0.03$ \\
219449 & K0III & $4853 \pm 130$ & 3.71 & 90 & $94.46 \pm 2.90$ & $0.45 \pm 0.02 $ & $2.260 \pm 0.126$ & $11.17 \pm 0.64$ \\
219828 & G0IV & $5929 \pm 90$ & 2.47 & 23 & $1.70 \pm 0.01$ & $0.08 \pm 0.01 $ & $0.203 \pm 0.006$ & $1.58 \pm 0.10$ \\
221287 & F6V & $6280 \pm 70$ & 0.04 & 27 & $2.07 \pm 0.02$ & $0.07 \pm 0.01 $ & $0.200 \pm 0.005$ & $1.19 \pm 0.05$ \\
222582 & G5V & $5636 \pm 50$ & 0.17 & 30 & $2.62 \pm 0.03$ & $0.18 \pm 0.01 $ & $0.263 \pm 0.005$ & $1.18 \pm 0.04$ \\
224693 & G0IV & $5929 \pm 90$ & 1.42 & 23 & $1.31 \pm 0.01$ & $0.01 \pm 0.01 $ & $0.178 \pm 0.006$ & $1.89 \pm 0.18$ \\
231701 & F5IV & $6562 \pm 150$ & 0.37 & 15 & $0.86 \pm 0.01$ & $0.31 \pm 0.02 $ & $0.118 \pm 0.005$ & $1.50 \pm 0.20$ \\
\enddata
\tablecomments{$N_{\rm PHOT}$ is the number of photometric data points available in the literature used for the spectral template fitting described in \S \ref{sec_sedfit}.}
\end{deluxetable}
\clearpage


\end{document}